\documentclass[aps,prb,twocolumn,floats]{revtex4}
\usepackage{graphics}
\usepackage{amssymb}
\usepackage{amsmath}
\begin{document}

\title{Heitler-London model for acceptor-acceptor interactions in doped semiconductors}
\author{Adam C. Durst$^1$, Kyle E. Castoria$^1$, and R. N. Bhatt$^2$}
\affiliation{$^1$Department of Physics and Astronomy, Hofstra University, Hempstead, NY 11549-1510}
\affiliation{$^2$Department of Electrical Engineering, Princeton University, Princeton, NJ 08544}
\date{August 29, 2017}

\begin{abstract}
The interactions between acceptors in semiconductors are often treated in qualitatively the same manner as those between donors.  Acceptor wave functions are taken to be approximately hydrogenic and the standard hydrogen molecule Heitler-London model is used to describe acceptor-acceptor interactions.  But due to valence band degeneracy and spin-orbit coupling, acceptor states can be far more complex than those of hydrogen atoms, which brings into question the validity of this approximation.  To address this issue, we develop an acceptor-acceptor Heitler-London model using single-acceptor wave functions of the form proposed by Baldereschi and Lipari, which more accurately capture the physics of the acceptor states.  We calculate the resulting acceptor-pair energy levels and find, in contrast to the two-level singlet-triplet splitting of the hydrogen molecule, a rich ten-level energy spectrum.  Our results, computed as a function of inter-acceptor distance and spin-orbit coupling strength, suggest that acceptor-acceptor interactions can be qualitatively different from donor-donor interactions, and should therefore be relevant to the control of two-qubit interactions in acceptor-based qubit implementations, as well as the magnetic properties of a variety of p-doped semiconductor systems.  Further insight is drawn by fitting numerical results to closed-form energy-level expressions obtained via an acceptor-acceptor Hubbard model.
\end{abstract}

\maketitle

\section{Introduction}
\label{sec:intro}
The donor-pair exchange interaction was studied several decades ago, when the main interest was magnetic susceptibility, magnetization, heat capacity and other thermodynamic properties of insulating n-doped semiconductors.  At that time, the stark difference between those materials and insulating spin glasses like Eu$_x$Sr$_{1-x}$S was very puzzling \cite{kum78}, and led to the understanding of highly disordered quantum antiferromagnets using a strong disorder renormalization group (SDRG) technique through the work of Bhatt and Lee \cite{bha81,bha82} as well as the one-dimensional counterpart by Dasgupta and Ma \cite{das80} which was put on rigorous analytical footing by Fisher \cite{fis94}.  This opened up a whole new set of RG techniques, which have been used for decades \cite{igl05,wes95,hym96,yan98}.

However, with the advent of quantum computation, dopant spins in silicon and other semiconductors became leading candidates for implementing qubits \cite{kan98,zwa13}, which demanded further scrutiny of the properties of such systems \cite{mor10,tyr12,sae13,pic14a,pic14b,kaw14,kim15,pic16a,pic16b,mi17}.  Controlling two qubit interactions is challenging and requires detailed knowledge of dopant-pair exchange, which has inspired renewed interest in this problem.  One of the issues for donor-based qubits in multi-valley semiconductors like Si, Ge, and AlAs is that the donor-donor interaction has a large oscillatory part as a function of inter-donor distance, and the oscillation occurs on atomic length scales, often requiring quite precise placement of dopant atoms, which can be problematic \cite{pic14a}.

For this reason, amongst others, there have been a number of recent proposals for acceptor-based qubits \cite{gol03,rus13,van14,aba16,sal16a,sal16b,aba17,van17}.  For acceptors, there aren't multiple valleys, so exchange is monotonic, and only involves the larger impurity (Bohr) radius.  This may make these systems more attractive under certain circumstances.  However, due to the anisotropic, degenerate nature of the valence band in most semiconductors, the behavior of bound acceptor holes is often more complex than that of donor electrons.  In this paper, we study the acceptor-pair exchange interaction in order to see how the more complex nature of acceptors affects exchange couplings in p-doped semiconductors.

At low densities and temperatures, donor electrons reside in bound states about positively charged donor ions and acceptor holes lie in bound states about negatively charged acceptor ions.  This situation is analogous to that of the electron bound to a proton in a hydrogen atom, except here the electron or hole dispersion is inherited from that of the host semiconductor.  Further, theoretical analyses of experiments \cite{tho81,bha80,bha87} show that the dominant effect of finite density on the properties of isolated impurities is due to the nearest neighbor, and it therefore suffices to consider impurity-pair interactions to describe most properties (e.g.\ optical, single-particle states, magnetic).  It has been shown that for magnetic properties in particular, pair interactions are able to quantitatively describe experiments throughout the insulating phase for phosphorus donors in silicon \cite{bha80,bha87,bha86}.

For the purpose of treating dopant interactions, dispersion is often assumed parabolic and isotropic and donors and acceptors are viewed as hydrogen atoms fixed within the semiconductor lattice, modified only by the use of an effective mass and dielectric constant.  In this approximation, the dopant-pair system can be described by the standard hydrogen molecule Heitler-London model. \cite{hei27,sla63}  While it is known that the Heitler-London model is an approximation, and does not agree with the known exact result at asymptotically large distances \cite{her64} for hydrogenic donor-pairs, it is quite accurate for typical distances encountered in doped semiconductors ($\sim4$ to 8 Bohr radii).

Since the conduction band of most semiconductors is very nearly parabolic at small wave vectors, this effective hydrogen atom approximation is usually adequate for describing the behavior of bound donor electrons, but due to the anisotropic, degenerate structure of the valence band, it is typically inadequate for describing bound acceptor holes.  In order to better describe the nature of acceptor states in semiconductors, more precise models of the single acceptor system have been developed by Schechter \cite{sch62}, Mendelsohn and James \cite{men64}, and Baldereschi and Lipari \cite{bal73,bal74}.  Since acceptor wave functions are generally more complex than simple hydrogenic wave functions, the behavior of the acceptor-pair system should naturally be more complex than that of the hydrogen molecule.  It is therefore questionable whether the acceptor-pair system can be accurately described by the standard hydrogen molecule Heitler-London model.

Recently, Salfi {\it et al.\/} \cite{sal16a} employed spatially-resolved tunneling to measure the energy spectrum of interacting acceptors near the surface of silicon.  Their experimental results, interpreted via theoretical calculations performed within the Kohn-Luttinger framework \cite{kav04,cli08,yak10,pas14}, were indicative of a spectrum far richer than the singlet-triplet splitting of the hydrogen molecule.

In the present work, we make use of the spherical acceptor wave functions of Baldereschi and Lipari \cite{bal73} to construct a Heitler-London model specifically for the acceptor-pair system.  (Preliminary work was reported in Ref.~\onlinecite{dur96}.)  Our calculations are performed numerically, as a function of inter-acceptor separation and spin-orbit coupling strength.  Computed energy spectra are then fit to closed-form expressions derived for an acceptor-acceptor Hubbard model.  Results shed light on the physics of acceptor-acceptor interactions and should therefore be applicable to the understanding and control of interactions between acceptor-based qubits, as well as the study of the magnetic properties of a variety of p-doped semiconductor systems, ranging from boron-doped silicon \cite{roy86} to diluted magnetic semiconductors like Ga$_{1-x}$Mn$_x$As \cite{zar02,fie03,fie05a,fie05b}.

Past magnetic experimental data on acceptor-doped semiconductors (e.g.\ Si:B) have been addressed using an ad-hoc generalization of the donor-pair model, i.e.\ using an isotropic $J = 3/2$ Heisenberg interaction \cite{roy86}.  By studying the more complicated acceptor-pair spectrum, the present work sets the stage for implementing an RG scheme like Bhatt-Lee \cite{bha81,bha82} but for p-doped systems, thereby providing a more convincing calculation of the thermodynamic properties of p-doped semiconductors in the insulating regime.

In Sec.~\ref{sec:hydrogen}, we review the standard Heitler-London model for the hydrogen molecule.  In Sec.~\ref{sec:single}, we consider the nature of acceptor states in p-doped semiconductors and discuss the single acceptor model developed by Baldereschi and Lipari \cite{bal73}.  With this background material established, we develop an acceptor-pair Heitler-London model in Sec.~\ref{sec:acceptorpair} and discuss its numerical implementation in Sec.~\ref{sec:numerics}.  The resulting acceptor-acceptor interaction energies are presented as a function of inter-acceptor separation and spin-orbit coupling strength in Sec.~\ref{sec:results} and explained in Sec.~\ref{sec:explanation} via a fit to a generalized Hubbard model developed in the Appendix.  Conclusions are discussed in Sec.~\ref{sec:conclusions}.

\section{Review: Hydrogen Molecule Heitler-London Model}
\label{sec:hydrogen}
The hydrogen molecule, H$_{2}$, is a system of two protons ($A$ and $B$)
and two electrons (1 and 2) interacting via pairwise Coulomb potentials.
Since the mass of a proton is nearly 2000 times that of an electron,
the protons are assumed (via the adiabatic approximation) to be fixed
in space with respect to the electrons, separated by an inter-proton
distance $R$.  (Such an approximation is even better in semiconductors where
the donors/acceptors sit substitutionally on the lattice sites of the host
semiconductor.)  The hydrogen molecule Hamiltonian therefore takes the form
\begin{equation}
H = -\nabla^{2}_{1} - \nabla^{2}_{2} - \frac{2}{r_{1A}} - \frac{2}{r_{2A}}
- \frac{2}{r_{1B}} - \frac{2}{r_{2B}} + \frac{2}{r_{12}} + \frac{2}{R}
\label{eq:hydrogenH}
\end{equation}
where energy is in units of Rydbergs and distance is in units of
the Bohr radius, $a_B$.  From left to right, the terms represent the kinetic
energies of the electrons, the potentials between the electrons and
proton $A$, the potentials between the electrons and proton $B$, the
inter-electron repulsion, and the inter-proton repulsion. \cite{sla63}

In the limit of large $R$, the individual hydrogen atoms are nearly
independent, with one electron bound to site $A$ (proton $A$) and the other
bound to site $B$ (proton $B$).  The essence of the Heitler-London \cite{hei27} model is to use the
large $R$ product states (i.e.\ $A(1)B(2)$ and $A(2)B(1)$) as a basis for the
evaluation of the hydrogen molecule Hamiltonian.  Notice that the model
explicitly excludes the possibility that two electrons will be bound to
the same site.  Due to the effects of inter-electron repulsion, this is
a good approximation.  (Hydrogen molecules are bonded covalently, not
ionically.)

\begin{figure}
\centerline{\resizebox{1.5in}{!}{\includegraphics{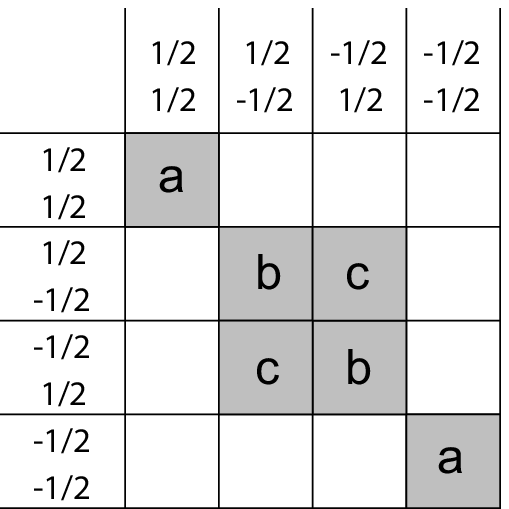}}}
\caption{Hydrogen molecule Heitler-London submatrix structure for both the H and S matrices.  Shaded elements are nonzero.
Equal elements share the same label.}
\label{fig:H2matelem}
\end{figure}

The basic formulation of the Heitler-London method \cite{hei27,sla63} is as follows.
Let $A_{\alpha}$ and $B_{\beta}$ represent the hydrogen atom ground state
wave functions about atom $A$ and atom $B$ respectively where $\alpha$ and
$\beta$ are electron spin indices ($\alpha,\beta = \pm \tfrac{1}{2}$).
Thus we have a total of four single atom orbitals:
$A_{\uparrow}$, $A_{\downarrow}$, $B_{\uparrow}$, and $B_{\downarrow}$.
Product states can be formed by choosing any two of these four orbitals.
Therefore, there are $4!/2!2! = 6$ possible product states of this sort.
However, the states $A_{\uparrow}A_{\downarrow}$ and $B_{\uparrow}B_{\downarrow}$
are excluded since they correspond to cases in which two electrons are bound
to a single site.  This leaves four basis states:
$A_{\uparrow}B_{\uparrow}$, $A_{\uparrow}B_{\downarrow}$,
$A_{\downarrow}B_{\uparrow}$, and $A_{\downarrow}B_{\downarrow}$,
each of which must be antisymmetric under exchange of electron 1 for electron 2
in order to satisfy the constraints of the Pauli exclusion principle.  To
ensure antisymmetry, we write each basis state as a Slater determinant
of the form
\begin{equation}
A_{\alpha} B_{\beta} = \left|
\begin{array}{cc} A_{\alpha}(1) & A_{\alpha}(2) \\
B_{\beta}(1) & B_{\beta}(2) \end{array} \right| .
\label{eq:slaterH2}
\end{equation}
Since these basis states are not necessarily orthogonal, we can form
a $4 \times 4$ overlap matrix with elements
$\langle A_{\alpha^{\prime}} B_{\beta^{\prime}} | A_{\alpha} B_{\beta} \rangle$.
Similarly, the hydrogen molecule Hamiltonian can be evaluated in this
basis to form a $4 \times 4$ Hamiltonian matrix with elements
$\langle A_{\alpha^{\prime}} B_{\beta^{\prime}} | H | A_{\alpha} B_{\beta} \rangle$.
Although each of these matrices has 16 components, the number of distinct
matrix elements is significantly reduced by taking advantage of symmetry.
Since the hydrogen molecule has inherent cylindrical symmetry about the
axis joining the protons ($z$-axis), it can be shown \cite{sla63} that the
$z$-component of electron spin is a conserved quantity.  Thus, only states
with the same total $S_{z}$ can couple to form nonzero matrix elements.
Since $S^{\rm tot}_{z}=1$ for the state $A_{\uparrow}B_{\uparrow}$, $S^{\rm tot}_{z}=0$ for
$A_{\uparrow}B_{\downarrow}$ and $A_{\downarrow}B_{\uparrow}$, and $S^{\rm tot}_{z}=-1$
for $A_{\downarrow}B_{\downarrow}$, the $H$ and $S$ matrices can each be
reduced to two $1 \times 1$ submatrices and one $2 \times 2$
submatrix.  As a result, the matrices take the form shown in
Fig.~\ref{fig:H2matelem}, where the shaded elements are nonzero and equal elements share the same label.  These equalities are ensured by the symmetries of the system (up-down invariance and symmetry upon swapping the $A$ and $B$ site labels).  (Of course, for this simple hydrogen molecule case, the $H$ and $S$ matrices can be trivially diagonalized by transforming to a basis of $S^{\rm tot}$ eigenstates.  But since spin-orbit coupling prevents an analogous transformation for the acceptor-acceptor case considered in Sec.~\ref{sec:acceptorpair}, we proceed by using the product basis, which, while more cumbersome here, is more readily generalized to the acceptor-acceptor case.)

Absent spin-orbit interactions, each single-particle state can be written as the product of a spatial wave function and a spin wave function (i.e.\ $A_{\uparrow}(1) = A(1)\chi_{\uparrow}(1)$ and
$A_{\downarrow}(1) = A(1)\chi_{\downarrow}(1)$).  Plugging states of this form into Eq.~(\ref{eq:slaterH2}) yields
\begin{eqnarray}
|A_{\alpha} B_{\beta}\rangle &=& |A(1) B(2)\rangle |\chi_{\alpha}(1)
\chi_{\beta}(2) \rangle \nonumber\\
&-&|A(2) B(1)\rangle |\chi_{\alpha}(2)\chi_{\beta}(1) \rangle
\label{eq:H2halfket}
\end{eqnarray}
where each of the two terms on the right-hand-side is what we shall refer to as a {\it half-ket}.  Evaluation of overlap matrix elements using these states is straightforward and we find that
\begin{eqnarray}
\langle A_{\alpha'} B_{\beta'} |A_{\alpha} B_{\beta} \rangle &=& \langle A(1)|A(1) \rangle ^2  \times  \delta_{\alpha' , \alpha} \delta_{\beta' , \beta} \nonumber\\
&-&  \langle A(1)|B(1) \rangle ^2 \times \delta_{\alpha' , \beta} \delta_{\beta' , \alpha} .
\label{eq:H2overlap}
\end{eqnarray}
Note that this expression imposes the matrix element structure depicted in Fig.~\ref{fig:H2matelem} and furthermore requires that $a_S = b_S + c_S$, where our element naming convention is that of the figure and the $S$ subscript designates elements of the $S$ matrix.

\begin{figure}
\centerline{\resizebox{3.25in}{!}{\includegraphics{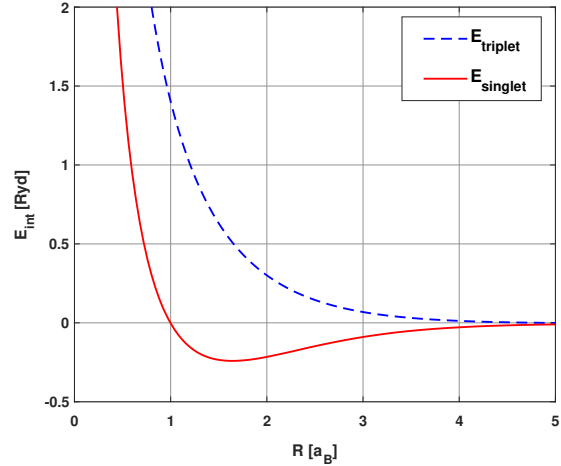}}}
\caption{Hydrogen molecule Heitler-London energy spectrum as a function of inter-proton distance $R$. $E_\text{triplet}$ is the three-fold degenerate, $S^{\rm tot}_z = 1$, excited state. $E_\text{singlet}$ is the nondegenerate, $S^{\rm tot}_z = 0$, ground state.}
\label{fig:hydrogen}
\end{figure}

The next step is to calculate the Hamiltonian matrix elements.  The process of doing so is simplified by rewriting Eq.~(\ref{eq:hydrogenH}) in terms of the single atom Hamiltonians.  Since electron 1 and electron 2 can each be bound to proton $A$ or proton $B$, there are four such Hamiltonians
\begin{eqnarray}
H_{1A} &=& - \nabla_{1}^{2} - \frac{2}{r_{1A}} \nonumber \\
H_{1B} &=& - \nabla_{1}^{2} - \frac{2}{r_{1B}} \nonumber \\
H_{2A} &=& - \nabla_{2}^{2} - \frac{2}{r_{2A}} \nonumber \\
H_{2B} &=& - \nabla_{2}^{2} - \frac{2}{r_{2B}} \nonumber \\
\label{eq:foursingleH2}
\end{eqnarray}
where, for example, $H_{1A}$ describes the single atom system where
electron 1 is bound to proton $A$.  Note that
$H_{1A}-2/r_{1B} = H_{1B}-2/r_{1A}$ and
$H_{2A}-2/r_{2B} = H_{2B}-2/r_{2A}$.  Therefore, plugging
Eq.~(\ref{eq:foursingleH2}) into Eq.~(\ref{eq:hydrogenH})
we find that the Hamiltonian can be expressed as
\begin{equation}
H = \left\{ \begin{array}{c} H_{1A} - \frac{2}{r_{1B}} \\
H_{1B} - \frac{2}{r_{1A}} \end{array} \right\}
+ \left\{ \begin{array}{c} H_{2B} - \frac{2}{r_{2A}} \\
H_{2A} - \frac{2}{r_{2B}} \end{array} \right\}
+ \frac{2}{r_{12}} + \frac{2}{R}
\label{eq:hydrogenH2}
\end{equation}
where the top and bottom components of the curly brackets are equal
to each other and the notation is intended to indicate that either the
top or bottom expressions can be used.  Since each of the single atom
wave functions are the ground state eigenfunctions of one of the hydrogen
Hamiltonians in Eq.~(\ref{eq:foursingleH2}), we know that
\begin{eqnarray}
H_{1A} | A_{\alpha}(1) \rangle &=& E_{0} | A_{\alpha}(1) \rangle \nonumber \\
H_{2A} | A_{\alpha}(2) \rangle &=& E_{0} | A_{\alpha}(2) \rangle \nonumber \\
H_{1B} | B_{\beta}(1) \rangle &=& E_{0} | B_{\beta}(1) \rangle \nonumber \\
H_{2B} | B_{\beta}(2) \rangle &=& E_{0} | B_{\beta}(2) \rangle
\label{eq:E0eigenH2}
\end{eqnarray}
where $E_{0}$ is the ground state energy of a hydrogen atom, equal to $-1$ in our Rydberg units.
Recalling that the basis kets have the Slater determinant form
$|A_{\alpha}B_{\beta}\rangle \equiv |A_{\alpha}(1)B_{\beta}(2)\rangle
- |A_{\alpha}(2)B_{\beta}(1)\rangle$
and making use of the above, we find that
\begin{eqnarray}
H | A_
{\alpha} B_{\beta} \rangle &=& \left( 2 E_{0} + 2/R
+ 2/r_{12} \right) | A_{\alpha} B_{\beta} \rangle \nonumber \\
&-& \left( 2/r_{1B} + 2/r_{2A} \right)
| A_{\alpha}(1) B_{\beta}(2) \rangle \nonumber \\
&+& \left( 2/r_{1A} + 2/r_{2B} \right)
| A_{\alpha}(2) B_{\beta}(1) \rangle .
\label{eq:HketH2}
\end{eqnarray}
Multiplying on the left by $\langle A_{\alpha'} B_{\beta'} |$ and taking the inner product yields the general form of a Hamiltonian matrix element. After reassigning electron indices and simplifying, Hamiltonian matrix elements take the form
\begin{eqnarray}
\lefteqn{\langle A_{\alpha'}B_{\beta'} |H|A_{\alpha}B_{\beta} \rangle =
\left( 2 E_{0} + 2/R \right) \langle A_{\alpha'} B_{\beta'} | A_{\alpha} B_{\beta} \rangle} \;\;\;\;\;\;\;\;\;\; \nonumber\\
&+& 2 \langle A(1)B(2)|\frac{2}{r_{12}} |A(1)B(2) \rangle
\delta_{\alpha' , \alpha} \delta_{\beta' , \beta} \nonumber\\
&-& 2 \langle A(2)B(1)|\frac{2}{r_{12}} |A(1)B(2) \rangle
\delta_{\beta' , \alpha} \delta_{\alpha' , \beta} \nonumber\\
&+& 4 \langle A(1)|\frac{2}{r_{1B}} |A(1) \rangle \langle B(2)|B(2) \rangle \delta_{\alpha' , \alpha} \delta_{\beta' , \beta} \nonumber \\
&-& 4 \langle B(1)|\frac{2}{r_{1B}} |A(1) \rangle \langle A(2)|B(2) \rangle \delta_{\beta' , \alpha} \delta_{\alpha' , \beta} . \nonumber\\
\label{eq:HbraketH2}
\end{eqnarray}
This expression imposes upon the Hamiltonian matrix the matrix element structure depicted in Fig.~\ref{fig:H2matelem} and requires that $a_H = b_H + c_H$, just as was the case for the overlap matrix.  Solving the secular equation, $\mbox{det} |\mathbf{H} - E\mathbf{S}|=0$, for each of the submatrices, reveals that the four energy levels take the form
\begin{eqnarray}
E_1 &=& \frac{a_{H}}{a_{S}}\nonumber\\
E_2 &=& \frac{b_{H}+c_{H}}{b_{S}+c_{S}} = \frac{a_{H}}{a_{S}}\nonumber\\
E_3 &=& \frac{b_{H}-c_{H}}{b_{S}-c_{S}}\nonumber\\
E_4 &=& \frac{a_{H}}{a_{S}} .
\label{eq:H2energies}
\end{eqnarray}
And since $a=b+c$ for both the $H$-matrix and the $S$-matrix, we see that $E_1 = E_2 = E_4$ is the three-fold degenerate triplet while $E_3$ is the nondegenerate singlet.  Subtracting $2E_0$ (the independent atom result) from each yields the triplet and singlet interaction energies, which we plot as a function of inter-proton distance $R$, in Fig.~\ref{fig:hydrogen}.

This well known result \cite{sla63} provides a baseline for the acceptor-acceptor generalization that will be derived in Sec.~\ref{sec:acceptorpair}.  To obtain a model for acceptor-acceptor interactions in doped semiconductors, this Heitler-London technique will be applied to more complex acceptor wave functions in place of the simple hydrogenic wave functions
used here.  As will be shown in the following section, unlike the hydrogenic wave functions, the single acceptor wave functions cannot be written as products of spatial wave functions and spin wave functions.  As a result, the generalized Heitler-London analysis will be more complex, resulting in ten energy levels rather than just two.

\section{Review: Single Acceptor Wave Functions}
\label{sec:single}
The primary difference between dopant states and hydrogenic states is
that dopants reside within semiconductor crystals while hydrogen atoms
live in free space.  As a result, the behavior of donor electrons and
acceptor holes is strongly influenced by the band structure of the
host semiconductor.  For most semiconductors, the low energy band structure
consists of a single minimum in the conduction band and a degenerate
maximum in the valence band.  Therefore, although donor states can be
modeled as hydrogenic states (with effective masses and dielectric
constants), acceptor states are more complex. \cite{bal73,bal74,koh57,and81}

Neglecting the effects of spin-orbit splitting, the typical valence
band maximum is 6-fold degenerate, including the 2-fold spin degeneracy.
Once the spin-orbit interaction is accounted for, some
of this degeneracy is lifted to reveal (at ${\bf k}=0$), a 4-fold degenerate top
band and a split-off 2-fold degenerate bottom band.
In nearly all semiconductors (except for Si) the splitting between the
top and bottom bands is large enough that the bottom band can be safely neglected.  In this investigation, we consider only
this large spin-orbit limit. \cite{sch62,koh57}

In this limit, the acceptor states must reflect the influence of a
4-fold degeneracy in the valence band.  Such a system is strongly
analogous to an atomic system in which the spin-orbit interaction
has been included. \cite{bal73}  In this analogy, each degenerate valence
band corresponds to a different atomic spin state.  Thus, the 4-fold
degenerate system in question can be modeled as an atomic system with
spin $J=3/2$ such that the $J_{z} = \{3/2, 1/2, -1/2, -3/2\}$ states
each correspond to the contribution of one of the degenerate bands.
In effect, the acceptor problem in the limit of strong spin-orbit splitting
is equivalent to the problem of a spin-3/2 particle in a Coulomb potential.
The Hamiltonian is given by
\begin{eqnarray}
H &=& (\gamma_{1} + \frac{5}{2} \gamma_{2}) \frac{p^{2}}{2m_{0}}
- \frac{\gamma_{2}}{m_{0}} \left( p_{x}^{2} J_{x}^{2} + p_{y}^{2} J_{y}^{2}
+ p_{z}^{2} J_{z}^{2} \right) - \frac{e^{2}}{\epsilon_{0} r} \nonumber \\
&& - \frac{2\gamma_{3}}{m_{0}} \Big( \{p_{x},p_{y}\} \{J_{x},J_{y}\}
+ \{p_{y},p_{z}\} \{J_{y},J_{z}\} \nonumber \\
&& \;\;\;\;\;\;\;\;\;\;\;\;\;\;\;\;\;\;\;\;\;\;\;\;\;\;\;\;\;\;\;\;\;\;\;\;\;
+ \{p_{z},p_{x}\} \{J_{z},J_{x}\} \Big)
\label{eq:accH}
\end{eqnarray}
where ${\bf p}$ is the hole momentum operator, ${\bf J}$ is the hole angular
momentum operator corresponding to spin-3/2, $\{a,b\} \equiv (ab + ba)/2$,
$\epsilon_{0}$ is the crystal dielectric constant, $m_{0}$ is the free
electron mass, and $\gamma_{1}$, $\gamma_{2}$, and $\gamma_{3}$ are the
Luttinger constants describing hole dispersion near ${\bf k}=0$. \cite{bal73,koh57}

Although the analogy with an atomic system is quite strong, the above
Hamiltonian does not have the full spherical symmetry of a true atomic
system.  Rather, it has only the cubic symmetry of a semiconductor crystal.
Using this cubic Hamiltonian, several models of the acceptor states have
been developed. \cite{sch62,men64}  However, it is possible to approximate
this cubic Hamiltonian by a Hamiltonian that has full spherical symmetry.  This
approximation procedure, in which the analogy to atomic systems is made
complete, has been developed by Baldereschi and Lipari \cite{bal73} and is
discussed in the following.

The first step in deriving the spherical model for acceptor states is
to separate the cubic Hamiltonian in Eq.~(\ref{eq:accH}) into
the sum of a term that has full spherical symmetry and a term that
has only cubic symmetry.  This is accomplished by writing the linear
and angular momentum operators in terms of irreducible spherical
tensors of rank two. \cite{bal73,edm57}
Doing so, Baldereschi and Lipari \cite{bal73} express Eq.~(\ref{eq:accH}) as
\begin{eqnarray}
H &=& - \nabla^{2} - \frac{2}{r} - \frac{\mu}{9 \hbar^{2}}
\left( P^{(2)} \cdot J^{(2)} \right) \nonumber \\
&& + \frac{\delta}{9 \hbar^{2}}
\Big( \left[ P^{(2)} \times J^{(2)} \right]^{(4)}_{4}
+ \frac{\sqrt{70}}{5} \left[ P^{(2)} \times J^{(2)} \right]^{(4)}_{0}
\nonumber \\
&& \;\;\;\;\;\;\;\;\;\;\;\;\;\;\;\;\;\;\;\;\;\;\;\;\;\;\;\;\;\;\;\;\;\;\;\;\;
+ \left[ P^{(2)} \times J^{(2)} \right]^{(4)}_{-4} \Big)
\label{eq:accH2}
\end{eqnarray}
where energies are given in units of the effective Rydberg,
${\rm Ryd} \equiv e^{4} m_{0} / 2 \hbar^{2} \epsilon_{0}^{2} \gamma_{1}$,
distances are given in units of the effective Bohr radius,
$a_B \equiv \hbar^{2} \epsilon_{0} \gamma_{1} / e^2 m_{0}$,
and the parameters $\mu \equiv (6 \gamma_{3} + 4 \gamma_{2}) / 5 \gamma_{1}$
and $\delta \equiv (\gamma_{3} - \gamma_{2}) / \gamma_{1}$
give the strength of the spherical spin-orbit interaction and the
cubic contribution respectively.  (Further discussion of this decomposition and definitions of the irreducible tensor notation can be found in Refs.~\onlinecite{bal73} and \onlinecite{edm57}.)  For nearly all semiconductors (with the notable exception of Si), the cubic parameter $\delta$ is much
smaller than $\mu$.  As a result, we can neglect the cubic contribution to
obtain the spherically symmetric Hamiltonian
\begin{equation}
H = - \nabla^{2} - \frac{2}{r} - \frac{\mu}{9 \hbar^{2}}
\left( P^{(2)} \cdot J^{(2)} \right) .
\label{eq:sphericalH}
\end{equation}

Neglecting the cubic terms, this system can be treated
just as one would treat a spherically symmetric atomic system.  Due to
the spherical symmetry of the Hamiltonian, the total angular momentum
${\bf F} = {\bf L} + {\bf J}$ is a conserved quantity where ${\bf L}$ is the orbital angular momentum
and ${\bf J}$ is the intrinsic particle spin ($=3/2$).  The ground state is
therefore characterized by $F=3/2$.  However, since the spin-orbit term
in the Hamiltonian couples states that differ in $L$ by 0 or 2, the most
general ground state wave function takes the form
\begin{eqnarray}
\Phi(S_{3/2})
&=& f_{0}(r) \left| L=0,J=3/2,F=3/2,F_{z} \right\rangle \nonumber \\
&+& g_{0}(r) \left| L=2,J=3/2,F=3/2,F_{z} \right\rangle
\label{eq:gndstate}
\end{eqnarray}
where $f_{0}$ and $g_{0}$ are radial functions and the $|L, J, F, F_{z} \rangle$
kets are eigenfunctions of $F^2$ and $F_z$ in the $L$-$J$ coupled scheme.  Taking the
matrix element of the spherical Hamiltonian in this ground state, Baldereschi and Lipari \cite{bal73} obtained
a set of two coupled second-order differential equations for the radial functions, $f_{0}$ and $g_{0}$.  In matrix form
\begin{equation}
H_0 |\Phi_0 \rangle = E |\Phi_0 \rangle \;\;\; {\rm where} \;\;\;
|\Phi_0 \rangle = \left[ \begin{array}{c} f_0(r) \\ g_0(r) \end{array} \right]
\;\;\; {\rm and} \nonumber
\end{equation}
\begin{equation}
H_0 = -\left[ \begin{array}{cc}
\frac{d^2}{dr^2}+\frac{2}{r} \frac{d}{dr}+\frac{2}{r}
& - \mu (\frac{d^2}{dr^2}+\frac{5}{r} \frac{d}{dr} + \frac{3}{r^2})  \\
 - \mu (\frac{d^2}{dr^2}+\frac{1}{r} \frac{d}{dr})
& \frac{d^2}{dr^2}+\frac{2}{r} \frac{d}{dr}- \frac{6}{r^2} +\frac{2}{r}
\end{array} \right] .
\label{eq:BLmatrixdiffeqs}
\end{equation}
Note that the minus sign in the definition of $H_0$, which does not appear in Ref.~\onlinecite{bal73}, defines $E$ to be the acceptor energy rather than the acceptor binding energy.

Baldereschi and Lipari \cite{bal73} employed a variational approach to estimate the ground state energy and radial functions.  Following their approach, we introduce trial radial functions of the form
\begin{equation}
f_{0}(r) = \sum_{i=1}^{21} A_{i} e^{-\alpha_{i}r^{2}} \;\;\;\;\;\;
g_{0}(r) = r \sum_{i=1}^{21} B_{i} e^{-\alpha_{i}r^{2}}
\label{eq:f0g0}
\end{equation}
where the $\alpha_{i}$ are constants chosen in geometric progression ($\alpha_{i+1} = g\alpha_i$ with $\alpha_1 = 10^{-2}$ and $\alpha_{21}=5 \times 10^5$) and
the $A_{i}$ and $B_{i}$ are variational parameters.  Evaluating the expectation value of $H_0$ given these trial radial functions, we define $E_0$ as a function of the 42 variational parameters
\begin{equation}
E_0 \equiv \langle H_0 \rangle = \frac{\langle \Phi_0 |H_0| \Phi_0 \rangle}{\langle \Phi_0 | \Phi_0 \rangle}
\label{eq:E0def}
\end{equation}
where
\begin{eqnarray}
\lefteqn{\langle \Phi_0 |H_0| \Phi_0 \rangle = 8\pi\sum_{ij} \Bigl[ (2\alpha_i^2 F_{ij4} - 3\alpha_i F_{ij2} + F_{ij1}) A_i A_j} \nonumber \\
&& + (2\alpha_j^2 F_{ij6} - 5\alpha_j F_{ij4} + F_{ij3} - 2F_{ij2} ) B_i B_j \nonumber \\
&& + \mu (8\alpha_j F_{ij3} - 2(\alpha_i^2 + \alpha_j^2) F_{ij5} - 4F_{ij1}) A_i B_j \Bigr]
\label{eq:E0num}
\end{eqnarray}
\begin{equation}
\langle \Phi_0 | \Phi_0 \rangle = 4\pi\sum_{ij} (F_{ij2}A_iA_j + F_{ij4}B_iB_j)
\label{eq:E0denom}
\end{equation}
and
\begin{equation}
F_{ijn} \equiv \int_0^ \infty  \! r^ne^{-(\alpha_i +\alpha_j) r^2} \, dr .
\label{eq:F}
\end{equation}
Note that this integral has simple closed-form solutions \cite{gra94} for all $i$, $j$, and $n$ that are encountered above.

We obtain optimal values for each of the variational parameters by taking derivatives of $E_0$ with respect to each and setting equal to zero.  Doing so yields
\begin{equation}
\frac{ \partial \langle \Phi_0 |H| \Phi_0 \rangle}{ \partial C_k}
= E_0 \frac{ \partial \langle \Phi_0 | \Phi_0 \rangle}{ \partial C_k}
\label{eq:linearequations}
\end{equation}
where $C_k$ is any one of the 42 variational parameters (the first 21 are the $A_i$ and the last 21 are the $B_i$).  Since both the numerator and denominator of Eq.~(\ref{eq:E0def}) are quadratic in each parameter, we obtain 42 linear equations for the 42 parameters, resulting in a $42 \times 42$ matrix equation of the form
\begin{equation}
{\bf P} {\bf c} = E_0 {\bf Q} {\bf c} .
\label{eq:42by42matrixequation}
\end{equation}
Solving this generalized eigenvalue problem yields 42 eigenvalues $E_{0\ell}$ and eigenvectors ${\bf c}_\ell$.  We seek only the minimum eigenvalue and its corresponding eigenvector.  The former yields the variational estimate of the ground state energy and the components of the latter yield the optimal values of the $A_i$ and $B_i$ variational parameters, and in turn, via Eq.~(\ref{eq:f0g0}), the optimal ground state radial functions.  Solutions reproduce the results of Ref.~\onlinecite{bal73} and depend on spin-orbit parameter $\mu$.  In this work, we compute a hundred such solutions, for $\mu$ ranging from 0 to 0.99.

\section{Acceptor-Pair Heitler-London Model}
\label{sec:acceptorpair}
The acceptor-pair system is analogous to that of the
hydrogen molecule.  It consists of two negatively charged acceptor
ions ($A$ and $B$) and two positively charged holes (1 and 2) interacting
via Coulomb potentials.  In the hydrogen molecule case, we assumed
that the protons were fixed with respect to the electrons because
the electron mass is so much less than the proton mass.  In this
acceptor-pair case, the acceptor ions really are fixed (barring
lattice vibrations) since they are physically embedded in the crystal
lattice.  The distribution of possible inter-acceptor distances, $R$,
is set by the dopant concentration of the semiconductor.  (Thus, overall
properties are approximated by a weighted average over the $R$-distribution.)
The main difference, of course, is that instead of interacting
hydrogen atoms, we have interacting acceptors.  Thus, representing the single-acceptor Hamiltonian via the spherical approximation of Baldereschi and Lipari \cite{bal73}, as discussed in Sec.~\ref{sec:single}, the acceptor-pair Hamiltonian takes the form
\begin{eqnarray}
H = \!\!\!\! && -\nabla_{1}^{2} - \nabla_{2}^{2} - \frac{2}{r_{1A}} - \frac{2}{r_{2A}}
- \frac{2}{r_{1B}} - \frac{2}{r_{2B}} + \frac{2}{r_{12}} + \frac{2}{R}
\nonumber \\
&& - \frac{\mu}{9 \hbar^{2}} \left( P^{(2)} \cdot J^{(2)} \right)_{1}
- \frac{\mu}{9 \hbar^{2}} \left( P^{(2)} \cdot J^{(2)} \right)_{2}
\label{eq:accpairH}
\end{eqnarray}
where energy is in units of effective Rydbergs and distance
is in units of effective Bohr radii.  From left to right,
the terms correspond to: the kinetic energy of the holes, the potential
between the holes and site $A$, the potential between the holes and site $B$,
the inter-hole repulsion, the inter-acceptor ion repulsion, and
spin-orbit terms for each of the holes.  Except for the
spin-orbit terms, this is exactly the hydrogen molecule Hamiltonian.

As prescribed by the Heitler-London approach, we take as basis
states the large-$R$-limit product states (orbital products where
there is one hole on each site).  Rather than using hydrogenic orbitals,
as in the hydrogen molecule case reviewed in Sec.~\ref{sec:hydrogen}, we now form product states
out of single acceptor orbitals, the Baldereschi-Lipari ground state
wave functions of the form given in Eq.~(\ref{eq:gndstate}).  In the hydrogen molecule case, since the hydrogen atom
wave functions were 2-fold degenerate in spin ($S_{z}=\pm 1/2$) and
each of the two sites had its own set of orbitals, there were a total of 4
orbitals to deal with.  However, in the acceptor-pair calculation, the
single acceptor wave functions are 4-fold degenerate in total angular momentum
($F_z= L_z + J_z =\pm 3/2,\pm1/2$).  Thus, between the two sites, there are a total of 8 single-acceptor orbitals.  We define these orbitals to be $A_{\alpha}$
and $B_{\beta}$ where $\alpha$ and $\beta$ can each assume the values
$\pm 3/2,\pm 1/2$.
Since product states are formed by choosing any 2 of these 8 orbitals,
there are $8!/2!(8-2)! = 28$ total product states.  However, $4!/2!2! = 6$
of these states have both holes on site $A$ and $4!/2!2! = 6$ of them have
both holes on site $B$.  Since the Heitler-London model excludes such
configurations, we are left with 16 suitable basis states of the form
$A_{\alpha}B_{\beta}$.  Since the holes are still fermions (spin 3/2), antisymmetry under exchange must again be enforced by expressing basis states as Slater determinants, via
Eq.~(\ref{eq:slaterH2}).  With 16 basis states, there are 256 overlap
matrix elements and 256 Hamiltonian matrix elements to compute.
Fortunately, we can make use of symmetries to reduce the total number
of distinct nonzero matrix elements.

\begin{figure}
\centerline{\resizebox{3.25in}{!}{\includegraphics{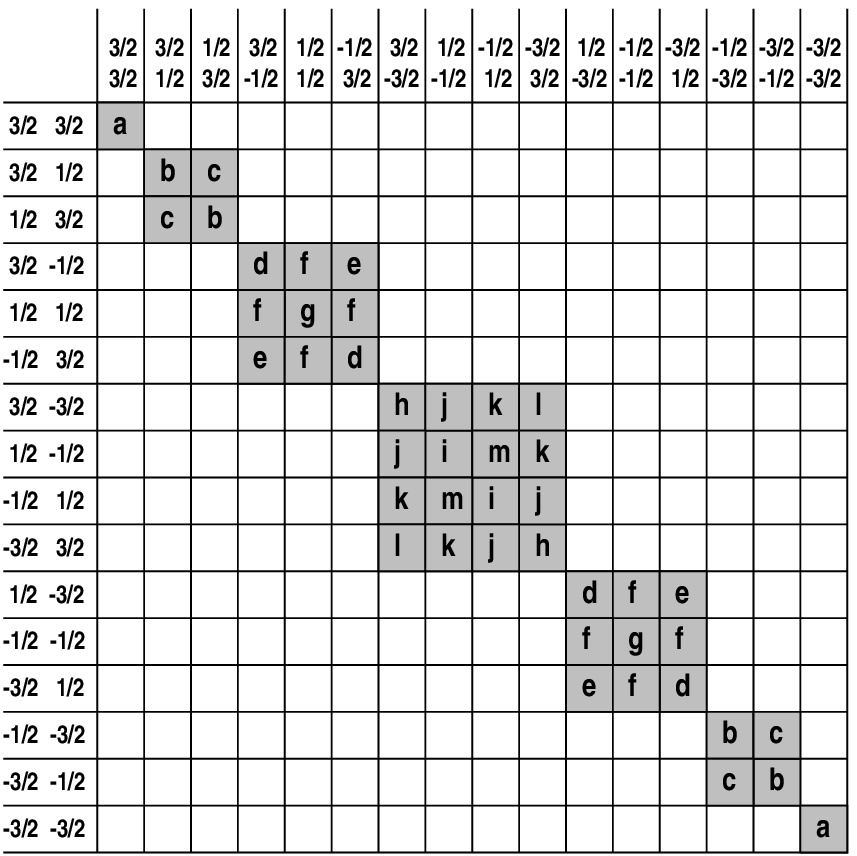}}}
\caption{Acceptor-pair Heitler-London submatrix structure.  Shaded elements are nonzero.  Equal elements share the same label.}
\label{fig:accmatelem}
\end{figure}

The acceptor-pair system has cylindrical symmetry about the line that
joins the two acceptor sites, which we are free to call the $z$-axis.  Therefore,
if we choose this to be the axis of quantization, the total angular
momentum along the $z$-axis must be a constant of the motion.  Hence,
$F_z^{\rm tot} \equiv F_{zA}+F_{zB}$ is a conserved quantity.  As a result, only basis states
with the same $F_z^{\rm tot}$ can couple to each other.  This reduces our
original $16 \times 16$ matrices to a group of submatrices: two $1 \times 1$
submatrices (for $F_z^{\rm tot} = \pm 3$), two $2 \times 2$ submatrices
(for $F_z^{\rm tot} = \pm 2$), two $3 \times 3$ submatrices (for
$F_z^{\rm tot} = \pm 1$), and one $4 \times 4$ submatrix (for $F_z^{\rm tot} = 0$).
Graphically, the matrices now take the form shown in Fig.~\ref{fig:accmatelem}
where the shaded elements are nonzero.

While there remain 44 elements that can have nonzero values, we can
utilize several other symmetries of the system to reduce the number of
distinct matrix elements even further.  First of all, notice that this
acceptor-pair system has up-down symmetry between the half
of the system containing site $A$ and the half containing site $B$.  The
consequences of this are two-fold.  First, this indicates that there is
nothing in the system to distinguish between positive $z$ and negative $z$.
Thus, matrix elements that transform into each other upon exchange of $F_{z}$
for $-F_{z}$ must be equal.  This reduces the number of distinct
elements to 24.  Second, the system must be invariant upon exchange of site $A$
for site $B$.  Thus, any matrix elements that transform into each other
when $A$ and $B$ are swapped must be equal.  This, in conjunction with the hermiticity of the matrices, brings the maximum number of
distinct nonzero matrix elements down to 13.  The arrangement of these
elements can be seen in Fig.~\ref{fig:accmatelem}.
By computing the 13 matrix elements and
solving the resulting secular equation, the energy levels of the acceptor-pair
system can be determined.

In the hydrogen molecule case, much can be
learned about that system by separating the wave functions into products of
spatial wave functions and spin wave functions in order to form spatial and spin
eigenfunctions.  However, such a procedure is not possible in this case
because the Baldereschi-Lipari single acceptor wave functions are not
separable.  This is clear from the form of these wave functions as presented
in Eq.~(\ref{eq:gndstate}).

Just as we simplified the hydrogen molecule Hamiltonian by taking advantage of the form of the four proton-electron Hamiltonians in Eq.~(\ref{eq:foursingleH2}), we can simplify the acceptor-pair Hamiltonian by taking advantage of the form of the four acceptor-hole Hamiltonians:
\begin{eqnarray}
H_{1A} &=& - \nabla_{1}^{2} - \frac{2}{r_{1A}} - \frac{\mu}{9 \hbar^{2}}
\left( P^{(2)} \cdot J^{(2)} \right)_{1} \nonumber \\
H_{1B} &=& - \nabla_{1}^{2} - \frac{2}{r_{1B}} - \frac{\mu}{9 \hbar^{2}}
\left( P^{(2)} \cdot J^{(2)} \right)_{1} \nonumber \\
H_{2A} &=& - \nabla_{2}^{2} - \frac{2}{r_{2A}} - \frac{\mu}{9 \hbar^{2}}
\left( P^{(2)} \cdot J^{(2)} \right)_{2} \nonumber \\
H_{2B} &=& - \nabla_{2}^{2} - \frac{2}{r_{2B}} - \frac{\mu}{9 \hbar^{2}}
\left( P^{(2)} \cdot J^{(2)} \right)_{2}
\label{eq:foursingleaccH}
\end{eqnarray}
Since these expressions are just those of Eq.~(\ref{eq:foursingleH2}) with the addition of a spin-orbit coupling term, it remains the case that $H_{1A}-2/r_{1B} = H_{1B}-2/r_{1A}$ and $H_{2A}-2/r_{2B} = H_{2B}-2/r_{2A}$.  Plugging into Eq.~(\ref{eq:accpairH}) yields
\begin{equation}
H = \left\{ \begin{array}{c} H_{1A} - \frac{2}{r_{1B}} \\
H_{1B} - \frac{2}{r_{1A}} \end{array} \right\}
+ \left\{ \begin{array}{c} H_{2B} - \frac{2}{r_{2A}} \\
H_{2A} - \frac{2}{r_{2B}} \end{array} \right\}
+ \frac{2}{r_{12}} + \frac{2}{R}
\label{eq:accpairH2}
\end{equation}
which is identical in form to Eq.~(\ref{eq:hydrogenH2}).  Like before, each of the single acceptor wave functions are the ground state eigenfunctions of one of the single acceptor Hamiltonians in Eq.~(\ref{eq:foursingleaccH}), therefore
\begin{eqnarray}
H_{1A} | A_{\alpha}(1) \rangle &=& E_{0} | A_{\alpha}(1) \rangle \nonumber \\
H_{2A} | A_{\alpha}(2) \rangle &=& E_{0} | A_{\alpha}(2) \rangle \nonumber \\
H_{1B} | B_{\beta}(1) \rangle &=& E_{0} | B_{\beta}(1) \rangle \nonumber \\
H_{2B} | B_{\beta}(2) \rangle &=& E_{0} | B_{\beta}(2) \rangle
\label{eq:E0eigen}
\end{eqnarray}
where $E_{0}$ is the Baldereschi-Lipari \cite{bal73} ground state energy of a single acceptor.  Again recalling that our basis kets have the Slater determinant form
$|A_{\alpha}B_{\beta}\rangle \equiv |A_{\alpha}(1)B_{\beta}(2)\rangle - |A_{\alpha}(2)B_{\beta}(1)\rangle$
and making use of the above, we find that
\begin{eqnarray}
H | A_{\alpha} B_{\beta} \rangle &=& \left( 2 E_{0} + 2/R
+ 2/r_{12} \right) | A_{\alpha} B_{\beta} \rangle \nonumber \\
&-& \left( 2/r_{1B} + 2/r_{2A} \right)
| A_{\alpha}(1) B_{\beta}(2) \rangle \nonumber \\
&+& \left( 2/r_{1A} + 2/r_{2B} \right)
| A_{\alpha}(2) B_{\beta}(1) \rangle .
\label{eq:Hket}
\end{eqnarray}
It is convenient to subtract out the constant, $2E_{0}+2/R$,
and rewrite the secular equation in the form
\begin{equation}
\mbox{det}\, | {\bf \Delta H} - \Delta E {\bf S} | = 0
\label{eq:deltasecular}
\end{equation}
where ${\bf \Delta H} \equiv {\bf H} - (2E_{0}+2/R)$ and
$\Delta E \equiv E - (2E_{0}+2/R)$.
Finally, hitting Eq.~(\ref{eq:Hket}) from the left with the basis bra
$\langle A_{\alpha^{\prime}}B_{\beta^{\prime}}|$ and noting that we
are free to interchange our labeling of hole 1 and hole 2, we obtain
the matrix elements of the acceptor-pair Hamiltonian:
\begin{eqnarray}
\langle A_{\alpha'} B_{\beta'} |\Delta H|A_{\alpha}B_{\beta} \rangle &=&
2\langle A_{\alpha'}(1)B_{\beta'}(2)|\frac{2}{r_{12}} |A(1)_{\alpha}B_{\beta}(2) \rangle
\nonumber \\
&-&2 \langle A_{\alpha'}(2)B_{\beta'}(1)|\frac{2}{r_{12}} |A(1)_{\alpha}B_{\beta}(2) \rangle \nonumber\\
&+&4 \langle A_{\alpha'}(1)B_{\beta'}(2)|\frac{2}{r_{1B}} |A_{\alpha}(1)B_{\beta}(2) \rangle \nonumber \\
&-&4 \langle A_{\alpha'}(2)B_{\beta'}(1)|\frac{2}{r_{1B}} |A_{\alpha}(1)B_{\beta}(2) \rangle \nonumber \\
\label{eq:DeltaHmatelem}
\end{eqnarray}
Similarly, the overlap matrix elements have the form
\begin{eqnarray}
\langle A_{\alpha'} B_{\beta'} |A_{\alpha}B_{\beta} \rangle &=&
2\langle A_{\alpha'}(1)B_{\beta'}(2)|A(1)_{\alpha}B_{\beta}(2) \rangle
\nonumber \\
&-&2 \langle A_{\alpha'}(2)B_{\beta'}(1)|A(1)_{\alpha}B_{\beta}(2) \rangle . \nonumber\\
\label{eq:overlap}
\end{eqnarray}

In order to evaluate matrix elements of ${\bf S}$ and ${\bf \Delta H}$ in the
$|A_{\alpha}B_{\beta}\rangle$ basis,
the basis states (kets) must be decomposed into sums and products of
orbital and spin wave functions.  First of all, we write the full-ket,
in terms of half-kets in the antisymmetric form,
$|A_{\alpha}B_{\beta}\rangle \equiv |A_{\alpha}(1)B_{\beta}(2)\rangle
- |A_{\alpha}(2)B_{\beta}(1)\rangle$.
Each of these half-kets is equal to the product of two Baldereschi-Lipari single acceptor wave functions (based on different sites).
\begin{eqnarray}
\label{eq:halfket}
\lefteqn{| A_{\alpha}(1) B_{\beta}(2) \rangle} \\
&&= \left[ f_{0}(r_{1A}) | L=0 , F_{z}=\alpha \rangle
+ g_{0}(r_{1A}) | L=2 , F_{z}=\alpha \rangle \right] \nonumber \\
&&\times \left[ f_{0}(r_{2B}) | L=0 , F_{z}=\beta \rangle
+ g_{0}(r_{2B}) | L=2 , F_{z}=\beta \rangle \right] \nonumber
\end{eqnarray}
Since each of the Baldereschi-Lipari wave functions has two terms, each half-ket is equal
to the sum of four terms that we call coupled-kets since the angular
momentum terms are given in the $L$-$J$ coupled scheme.  For example, the
third coupled-ket in the decomposition of Eq.~(\ref{eq:halfket}) is
\begin{equation}
| \mbox{Coup} \rangle = g_{0}(r_{1A}) f_{0}(r_{2B})
| L=2 , F_{z}=\alpha \rangle |L=0 , F_{z}=\beta \rangle .
\label{eq:coupket}
\end{equation}
If all the matrix elements that we needed to calculate had both a bra and
a ket based on the same acceptor site, then the spherical symmetry of
the system would allow us to evaluate matrix elements in the coupled
$L$-$J$ basis.  However, since some of the matrix elements have their bra
and ket based on different sites, spherical symmetry is broken and total
angular momentum $F$ is no longer a good quantum number.  As a result, the matrix elements must be evaluated
in the decoupled scheme where $L$ and $J$ are separable and the basic angular
momentum kets are of the form $|L L_{z}\rangle |J J_{z}\rangle$.
The terms in the coupled basis can be written
as sums of terms in the decoupled basis weighted by a series of
Clebsch-Gordan coefficients \cite{edm57,gri05}.  Specifically, the $0 \times 3/2$ Clebsch-Gordan
table (a trivial one) is used to decompose terms like
$|L=0, J=3/2, F=3/2, F_{z}=\alpha \rangle$ and the $2 \times 3/2$ table \cite{eur00}
is used to decompose terms like $|L=2, J=3/2, F=3/2, F_{z}=\beta \rangle$.
Using these tables, each $|L~F_z\rangle$ ket in Eq.~(\ref{eq:coupket}) can be decomposed via one of the following
\begin{eqnarray}
|0, \alpha \rangle &=& Y_0^0 | \alpha \rangle \nonumber\\
\left| 2, \tfrac{3}{2} \right\rangle &=& \sqrt{\tfrac{2}{5}} Y_2^2 \left| \tfrac{-1}{2} \right\rangle
- \sqrt{\tfrac{2}{5}} Y_2^1 \left| \tfrac{1}{2} \right\rangle
+ \sqrt{\tfrac{1}{5}} Y_2^0 \left| \tfrac{3}{2} \right\rangle\nonumber\\
\left| 2, \tfrac{1}{2} \right\rangle &=& \sqrt{\tfrac{2}{5}} Y_2^2 \left| \tfrac{-3}{2} \right\rangle
+ \sqrt{\tfrac{2}{5}} Y_{2}^{-1} \left| \tfrac{3}{2} \right\rangle
- \sqrt{\tfrac{1}{5}} Y_2^0 \left| \tfrac{1}{2} \right\rangle\nonumber\\
\left| 2, \tfrac{-1}{2} \right\rangle &=& \sqrt{\tfrac{2}{5}} Y_2^{-2} \left| \tfrac{3}{2} \right\rangle
+ \sqrt{\tfrac{2}{5}} Y_{2}^{1} \left| \tfrac{-3}{2} \right\rangle
- \sqrt{\tfrac{1}{5}} Y_2^0 \left| \tfrac{-1}{2} \right\rangle\nonumber\\
\left| 2, \tfrac{-3}{2} \right\rangle &=& \sqrt{\tfrac{2}{5}} Y_2^{-2} \left| \tfrac{1}{2} \right\rangle
- \sqrt{\tfrac{2}{5}} Y_{2}^{-1} \left| \tfrac{-1}{2} \right\rangle
+ \sqrt{\tfrac{1}{5}} Y_2^0 \left| \tfrac{-3}{2} \right\rangle\nonumber\\
\label{eq:clebschgordandecomp}
\end{eqnarray}
where the kets on the right are $J_z$ eigenstates and the $Y_l^m$ are spherical harmonics centered upon the acceptor site in question.

The halfkets in Eq.~(\ref{eq:DeltaHmatelem}) were each broken into 4 terms via Eq.~(\ref{eq:halfket}). Next Eq.~(\ref{eq:clebschgordandecomp}) broke that group of 4 terms into 16, and hit with a similar state on the left, that 16 squares to 256. Noting that Eq.~(\ref{eq:DeltaHmatelem}) itself has four terms and that Eq.~(\ref{eq:overlap}) has two more, the matrix elements for a given combination of spins then requires the evaluation of 1,536 terms, 512 of which contain 6D integrals and 1,024 of which contain two 3D integrals.

\begin{figure}
\centerline{\resizebox{3.25in}{!}{\includegraphics{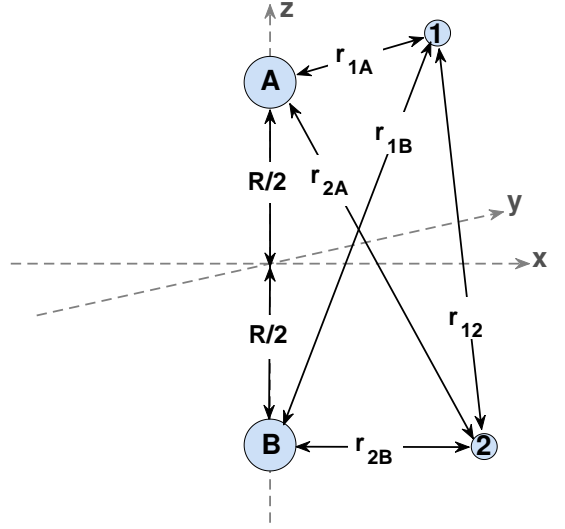}}}
\caption{Geometry of the acceptor-acceptor system.  Depicted are acceptors $A$ and $B$ and holes 1 and 2, as well as the distances between them, as defined in Eq.~(\ref{eq:lengths}).}
\label{fig:sketchI}
\end{figure}

It is simplest to evaluate these integrals in cylindrical coordinates ($\rho_1$, $z_1$, $\phi_1$ for hole 1 and $\rho_2$, $z_2$, $\phi_2$ for hole 2) where the $z$-axis joins the acceptor ions and the origin is located at the midpoint between them.  The relevant interaction distances, shown in Fig.~\ref{fig:sketchI}, are then expressed via
\begin{eqnarray}
r_{1A} &=& \sqrt{\rho_1^2 + (z_1 - R/2)^2}\nonumber\\
r_{1B} &=& \sqrt{\rho_1^2 + (z_1 + R/2)^2}\nonumber\\
r_{2A} &=& \sqrt{\rho_2^2 + (z_2 - R/2)^2}\nonumber\\
r_{2B} &=& \sqrt{\rho_2^2 + (z_2 + R/2)^2}\nonumber\\
r_{12} &=& \sqrt{\rho_1^2 + \rho_2^2 - 2\rho_1 \rho_2\cos(\phi_2 -\phi_1)+(z_2-z_1)^2}\nonumber\\
\label{eq:lengths}
\end{eqnarray}
where $r_{iJ}$ is the distance from hole $i$ to acceptor $J$, $r_{12}$ is the inter-hole distance, and $R$ is the inter-acceptor distance.

The 3D integrals that emerge from the terms that do not depend on $r_{12}$ have the form
\begin{equation}
I_3 = C_{1'} C_1 \int_0^\infty \!\!\!\!\!\!d\rho \int_{-\infty}^\infty \!\!\!\!\!\!dz\, X(r_{1B}) h_1(r_{1A}) h_2(r_{1J})
\int_{-\pi}^\pi \!\!\!\!\!\!d\phi\, Y_{l'}^{m'*} Y_l^m
\label{eq:I3}
\end{equation}
where the $X$-function can be either $1$ or $2/r$, the $h$-functions can be either $f_0(r)$ or $g_0(r)$, the $C$-constants are Clebsch-Gordan coefficients, and $J$ can denote either acceptor $A$ or acceptor $B$.  Since the spherical harmonics have the functional form $Y_l^m = y_l^m(z,\rho)e^{im\phi}$, the $\phi$ integral can be performed analytically
\begin{equation}
\int_{-\pi}^{\pi}  \! Y^{m'*}_{l'}Y^m_l \, d\phi = y^{m'}_{l'}y^m_l 2\pi \delta_{m',m}
\label{eq:phiintegral}
\end{equation}
leaving 2D integrals to be performed numerically.

The 6D integrals that emerge from the terms that do depend on $r_{12}$ have the form
\begin{eqnarray}
I_6 &=& C_{1'} C_1 C_{2'} C_2 \int_0^\infty \!\!\!\!\!\!d\rho_1 \int_{-\infty}^\infty \!\!\!\!\!\!dz_1\,
\int_0^\infty \!\!\!\!\!\!d\rho_2 \int_{-\infty}^\infty \!\!\!\!\!\!dz_2\, \nonumber \\
&& \times h_1(r_{1A}) h_2(r_{2B}) h_3(r_{1J}) h_4(r_{2\bar{J}}) \nonumber \\
&& \times \int_{-\pi}^\pi \!\!\!\!\!\!d\phi_1\, \int_{-\pi}^\pi \!\!\!\!\!\!d\phi_2\,
\frac{2}{r_{12}} Y_{l'_1}^{m'_1*} Y_{l_1}^{m_1} Y_{l'_2}^{m'_2*} Y_{l_2}^{m_2}
\label{eq:I6}
\end{eqnarray}
where the $h$-functions can be either $f_0(r)$ or $g_0(r)$, the $C$-constants are Clebsch-Gordan coefficients, $J$ denotes either acceptor $A$ or acceptor $B$, and $\bar{J}$ denotes the acceptor that $J$ does not.  Due to the functional form of the spherical harmonics, and the fact that $r_{12}$ depends only on the relative angle $\phi_2 - \phi_1$, one of the $\phi_i$ integrals can be performed analytically
\begin{eqnarray}
\int_{-\pi}^\pi \!\!\!\!\!\!d\phi_1\, \int_{-\pi}^\pi \!\!\!\!\!\!d\phi_2\,
\frac{2}{r_{12}} Y_{l'_1}^{m'_1*} Y_{l_1}^{m_1} Y_{l'_2}^{m'_2*} Y_{l_2}^{m_2} \nonumber \\
= y_{l'_1}^{m'_1} y_{l_1}^{m_1} y_{l'_2}^{m'_2} y_{l_2}^{m_2} 2\pi\delta_{m'_1+m'_2,m_1+m_2} \nonumber \\
\times \int_{-\pi}^\pi \!\!\!\!\!\!d\phi\, \frac{2}{r_{12}} \cos[(m_2 - m'_2)\phi]
\label{eq:phi1phi2integral}
\end{eqnarray}
where $\phi \equiv \phi_2 - \phi_1$.  The resulting 5D integrals are left to the numerics.

Thus, the task of computing the 26 unique elements of the ${\bf \Delta H}$ and ${\bf S}$ matrices becomes one of bookkeeping --- keeping track of all the terms born of Eqs.~(\ref{eq:DeltaHmatelem}) through (\ref{eq:clebschgordandecomp}) --- as well as numerical integration of the 2D and 5D integrals discussed above.  Details of these numerics are discussed in the following section.

\section{Numerics}
\label{sec:numerics}

Our numerical calculation proceeded in three stages.  First a Python script was used to organize each matrix element into a sum of labeled integrals.  By reassigning labels and permuting terms, it was determined that a total of 21 unique 2D integrals and 61 unique 5D integrals were required for any particular set of input parameter values.  The script assessed each matrix element, determined which terms had nonzero prefactors, and reorganized each of the surviving terms until it could be expressed in terms of the unique labeled integrals.  This bookkeeping step saved significant computation time by avoiding unnecessary repeated numerical computation of the same integrals.

Next, the labeled integrals were computed numerically.  For the 2D integrals, we used the Gaussian quadrature algorithm in Matlab.  But for the 5D integrals, equivalent methods proved prohibitively slow, so we used a Monte Carlo integration routine \cite{pre92} written in C++.  This was the most computationally expensive step of the calculation.

Finally, the integration results were loaded into Matlab, where they were used to reassemble the matrix elements of the Hamiltonian and overlap matrices.  The remaining linear algebra was performed there, where we solved the $16 \times 16$ secular equation via a generalized eigenvalue routine.  This yielded 16 energies, 6 of which were doubly degenerate by symmetry, and thereby produced the 10-level energy spectrum discussed in Sec.~\ref{sec:acceptorpair}.

The energy spectra depend on two parameters, the inter-acceptor distance $R$ and the spin-orbit parameter $\mu$.  Spectra were calculated over a grid of points in this two-dimensional parameter space, with inter-acceptor distance ranging from $R=0.04 a_B$ to $R=5 a_B$ in steps of $0.04 a_B$ and spin-orbit parameter ranging from $\mu=0$ to $\mu=0.99$ in steps of $0.01$.  Thus, in total, 12,500 spectra were calculated.

Note: For the hydrogen molecule, the Heitler-London model is known to be accurate in the large $R$ limit, but less so for $R \lesssim 2 a_B$.  We can quantify this by comparing the data in Fig.~\ref{fig:hydrogen} to the numerically exact results of Kolos and Wolniewicz \cite{kol65}.  Looking at the singlet-triplet splitting, we see that the hydrogen molecule Heitler-London model yields a 7.1\% error for $R=2a_B$, a 19.0\% error for $R=1.5a_B$, and a 39.6\% error for $R=1a_B$.  We therefore expect similar errors for our acceptor-pair calculation.  But despite this degradation in quantitative accuracy with decreasing $R$, the Heitler-London model does a fine job of illustrating the qualitative physics of the hydrogen molecule, even down to smaller $R$.  Our intent with this work is to do the same for the acceptor-acceptor problem, and it is in that spirit that we compute down to smaller $R$, despite expectations of decreasing quantitative accuracy with decreasing $R$.

To enable the collection of this large amount of data, the 5D Monte Carlo integration C++ code was parallelized and run on the Hofstra Big Data Cluster.  The cluster consists of 28 dual core CPUs that together can handle a total of 56 independent parallel calculations simultaneously.  In this configuration, each spectrum can be obtained in 36 seconds (down from roughly 11 minutes when run on a personal laptop) and a complete data set (12,500 spectra) required approximately 5 days of processing time (down from 94 days).  This reduction in computation time was essential as it allowed us to study in great detail how the energy spectra evolve as a function of both inter-acceptor distance and spin-orbit parameter.  The results of this analysis are presented in the following section.

\section{Results}
\label{sec:results}

The matrix elements of the Hamiltonian matrix, ${\bf H}$, and the overlap matrix, ${\bf S}$,
were calculated numerically for 100 values of the spin-orbit
constant, $\mu$, and 125 values of the inter-acceptor distance, $R$.
As discussed in Sec.~\ref{sec:acceptorpair}, the symmetries of the
acceptor-pair system limit the number of distinct nonzero matrix elements
of ${\bf H}$ and ${\bf S}$ to 13.  These 13 matrix elements are arranged in the block
diagonal form of Fig.~\ref{fig:accmatelem}, which consists of one $4 \times $4,
two $3 \times 3$, two $2 \times 2$, and two $1 \times 1$ submatrices.
Solving the secular equation [see Eq.~(\ref{eq:deltasecular})] for each submatrix yields 10 values of $\Delta E$, 6 of which are doubly degenerate.  The
total acceptor pair energies are then given by $E=\Delta E + 2/R + 2E_{0}$ where $E_0$ is the single acceptor energy.
Subtracting off the energy of the two independent acceptors yields the
interaction energy, $E_{\rm int} \equiv E - 2E_{0} = \Delta E + 2/R$.

\begin{figure}
\centerline{\resizebox{3.25in}{!}{\includegraphics{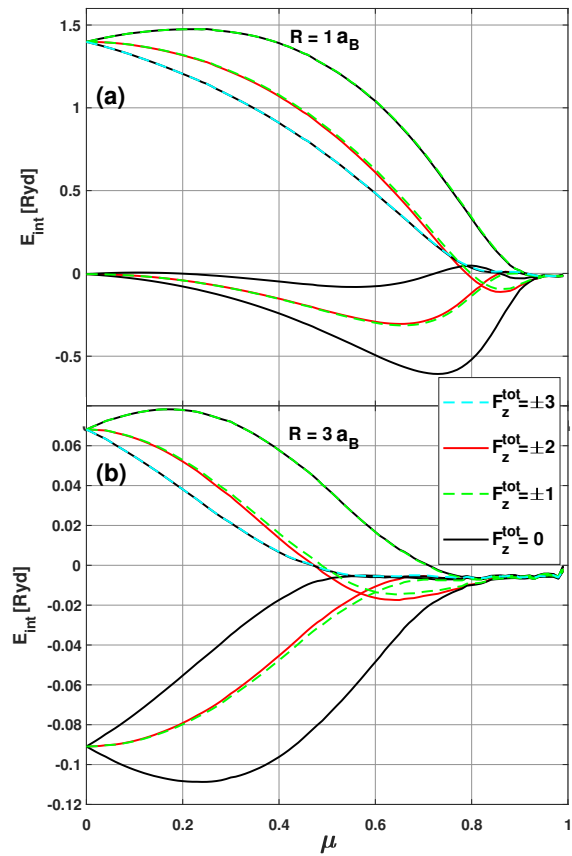}}}
\caption{Acceptor-pair Heitler-London energy spectra for $R=1$ and $R=3$
effective Bohr radii in panels (a) and (b) respectively.  Energy levels (in effective Rydbergs) are
plotted as a function of spin-orbit parameter $\mu$.  There are four nondegenerate $F_z^{\rm tot}=0$ levels (solid black), three 2-fold degenerate $F_z^{\rm tot}=\pm 1$ levels (dashed green), two 2-fold degenerate $F_z^{\rm tot}=\pm 2$ levels (solid red), and one 2-fold degenerate $F_z^{\rm tot}=\pm 3$ level (dashed blue).}
\label{fig:energyR}
\end{figure}

In Fig.~\ref{fig:energyR}, $E_{\rm int}$ is plotted as a function of spin-orbit parameter $\mu$ from $\mu=0$ to $\mu=0.99$, for two different values of inter-acceptor distance, $R=1a_B$ and $R=3a_B$, where $a_B$ is the effective Bohr radius.  Note that $\mu=0$ corresponds to the case of a hydrogenic acceptor wave function
while $\mu = 0.77$ is a realistic value for GaAs. \cite{bal73}

\begin{figure}
\centerline{\resizebox{3.25in}{!}{\includegraphics{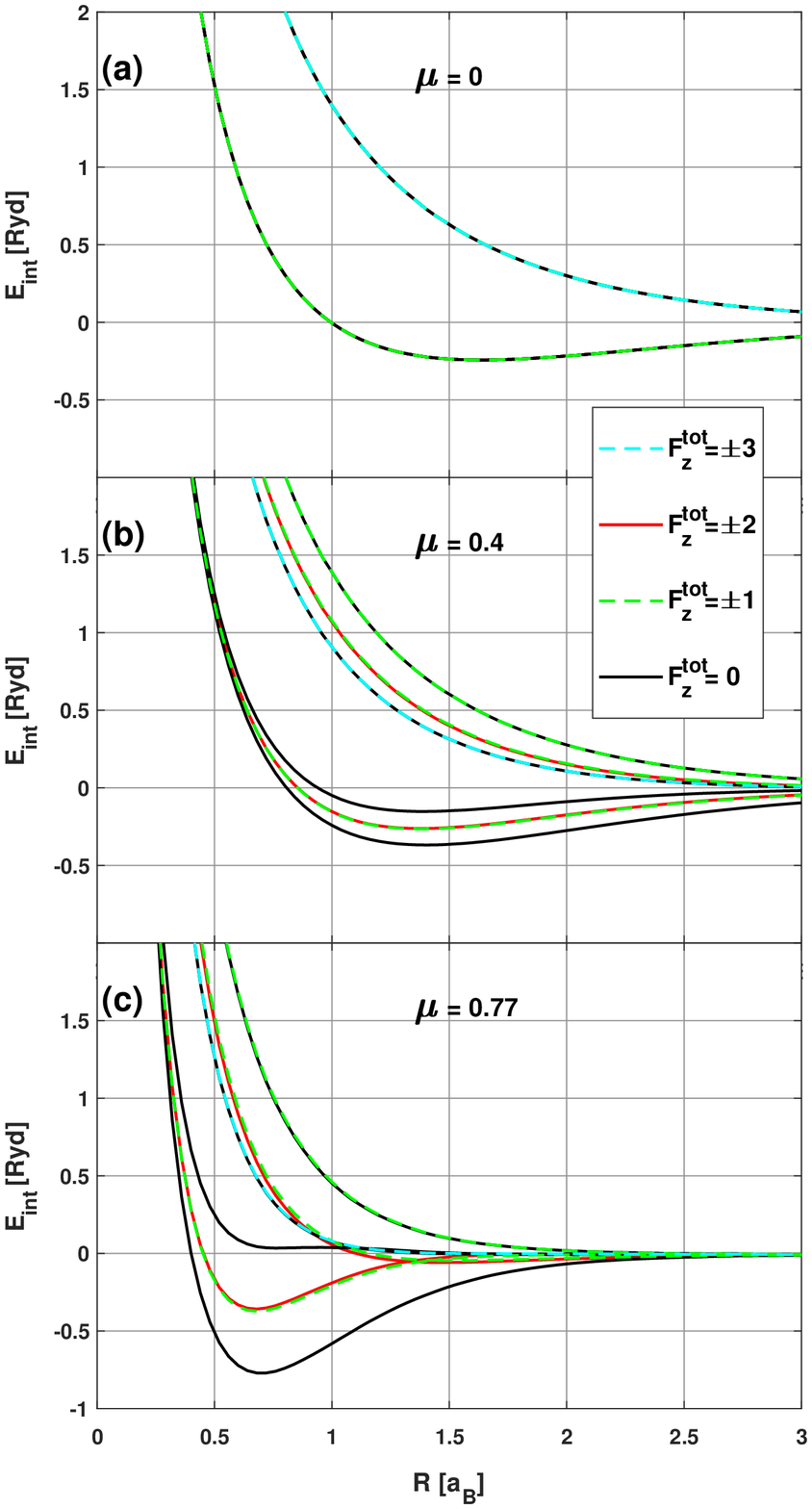}}}
\caption{Acceptor-pair Heitler-London energy spectra for (a) $\mu=0$, (b) $\mu=0.4$, and (c) $\mu=0.77$.  Energy levels (in effective Rydbergs) are
plotted as a function of inter-acceptor distance $R$ from $R=0$ to $R=3a_B$.  There are four nondegenerate $F_z^{\rm tot}=0$ levels (solid black), three 2-fold degenerate $F_z^{\rm tot}=\pm 1$ levels (dashed green), two 2-fold degenerate $F_z^{\rm tot}=\pm 2$ levels (solid red), and one 2-fold degenerate $F_z^{\rm tot}=\pm 3$ level (dashed blue).}
\label{fig:energyM1}
\end{figure}

\begin{figure}
\centerline{\resizebox{3.25in}{!}{\includegraphics{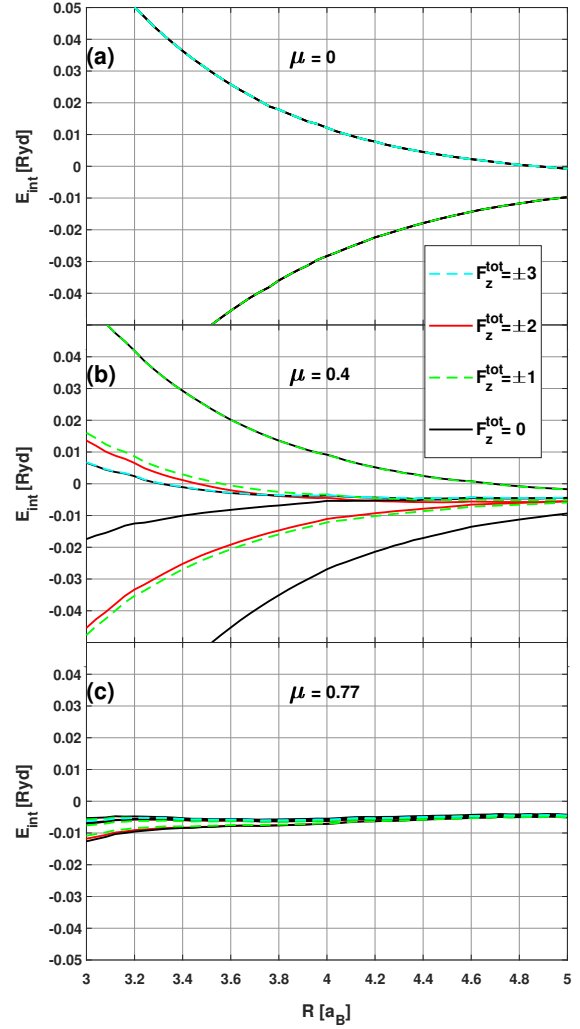}}}
\caption{Acceptor-pair Heitler-London energy spectra for (a) $\mu=0$, (b) $\mu=0.4$, and (c) $\mu=0.77$.  Energy levels (in effective Rydbergs) are
plotted as a function of inter-acceptor distance $R$ from $R=3a_B$ to $R=5a_B$.  There are four nondegenerate $F_z^{\rm tot}=0$ levels (solid black), three 2-fold degenerate $F_z^{\rm tot}=\pm 1$ levels (dashed green), two 2-fold degenerate $F_z^{\rm tot}=\pm 2$ levels (solid red), and one 2-fold degenerate $F_z^{\rm tot}=\pm 3$ level (dashed blue).  The noise in these plots reflects the statistical error in the numerical Monte Carlo computation of our 5D integrals.  It is more visible here than in Fig.~\ref{fig:energyM1} because we have zoomed in at a smaller energy scale. }
\label{fig:energyM2}
\end{figure}

For $\mu=0$, the hydrogenic limit, we have two distinct energy levels, a six-fold degenerate ground state and a ten-fold degenerate excited state.  This result is straightforward to understand.  In this limit, the single-acceptor Hamiltonian [Eq.~(\ref{eq:sphericalH})] becomes spin-independent and the $L=2$ term in the Baldereschi-Lipari wave function [Eq.~(\ref{eq:gndstate})] vanishes.  The remaining $L=0$ term is now just the product of a hydrogenic ground state spatial wave function and a pure spin state for $J=3/2$.  Since only $L=0$ states remain, $F$ labels become equivalent to $J$ labels.  The acceptor-pair Hamiltonian [Eq.~(\ref{eq:accpairH})] is also spin-independent, so $J^{\rm tot}$, and therefore $F^{\rm tot}$, are good quantum numbers.  The eigenstates of the (spin-independent) acceptor-pair Hamiltonian are product states of spatial wave functions and spin wave functions.  Though the Hamiltonian lacks spherical symmetry in space, the spin wave functions are independent of the spatial structure of the problem.  The spatial wave functions are precisely those of the hydrogen molecule, a ground state that is symmetric upon exchange and an excited state that is antisymmetric upon exchange.  Overall antisymmetry upon exchange requires that the ground state spatial wave function be multiplied by one of the six antisymmetric-upon-exchange spin wave functions (total spin 0 or 2) and the excited state spatial wave function be multiplied by one of the ten symmetric-upon-exchange spin wave functions (total spin 1 or 3).  The six-fold degenerate ground state is therefore labeled by $F^{\rm tot}=0$ ($F^{\rm tot}_{z}=0$) and
$F^{\rm tot}=2$ ($F^{\rm tot}_{z}=0, \pm 1, \pm 2$), and is the spin-3/2 analog of the
singlet.  The ten-fold degenerate excited state is labeled by $F^{\rm tot}=1$
($F^{\rm tot}_{z}=0, \pm 1$) and $F^{\rm tot}=3$ ($F^{\rm tot}_{z}=0, \pm 1, \pm 2, \pm 3$), and is the
spin-3/2 analogue of the triplet.  Thus, the energy splitting is precisely that of the familiar hydrogen molecule Heitler-London model. \cite{sla63}

But once $\mu$ is nonzero, the $L=2$ piece of the Baldereschi-Lipari wave function becomes nonzero, making $F$ and $J$ labels no longer equivalent, and the acceptor-pair Hamiltonian becomes spin-dependent, coupling spin to space, all of which means that total spin $F^{\rm tot}$ is no longer a good quantum number.  The energy levels split, first into six distinct levels and then further into ten levels.  Initially, the ground state splits into an $F^{\rm tot}_{z}=0$ level, an $F^{\rm tot}_{z}=\pm 1, \pm 2$ level, and another $F^{\rm tot}_{z}=0$ level. Note that the lowest energy level (the new ground state) is a nondegenerate ($F^{\rm tot}_z=0$) level.  The excited state initially splits into an $F^{\rm tot}_{z}=0, \pm 3$ level, an $F^{\rm tot}_{z}= \pm 1, \pm 2$ level, and an $F^{\rm tot}_{z}=0, \pm 1$ level.  For larger $\mu$, the 3-fold and 4-fold degenerate levels split further, resulting in a ten-level spectrum.  The high-energy levels stay separated from the low-energy levels for small enough $\mu$ and $R$.  However, increasing $\mu$ (for constant $R$) reduces this separation, and eventually the levels mix.  This mixing of levels happens at smaller values of $\mu$ for larger values of $R$, as can be seen in Fig.~\ref{fig:energyR}.  After the levels mix, they approach zero from below.

The plots in Fig.~\ref{fig:energyM1} and Fig.~\ref{fig:energyM2} show the energy levels as a function of $R$ at three different $\mu$ values. Fig.~\ref{fig:energyM1} shows the plots from $R=0$ to $R=3a_B$ on a large energy scale from $E_{\rm int} = -1\, {\rm Ryd}$ to $E_{\rm int} = 2\, {\rm Ryd}$.  Fig.~\ref{fig:energyM2} continues the plots to $R=5a_B$, but on a smaller scale between $E_{\rm int} = \pm 0.05\, {\rm Ryd}$.  The upper panels, where $\mu=0$, represent the hydrogenic limit. In this limit, the levels are identical to that of the hydrogen molecule (previously presented in Fig.~\ref{fig:hydrogen}), however the hydrogenic triplet is now 10-fold degenerate and the hydrogenic singlet is now 6-fold degenerate.  In the middle panels, where $\mu=0.4$, these two levels have split into six, of approximate degeneracy 1, 4, 1, 3, 4, and 3, as discussed above.  The closer view in Fig.~\ref{fig:energyM2}(b) shows the small splitting of the 3-fold and 4-fold levels, revealing that there are ten levels in all.  At this moderate spin-orbit coupling strength, the high-energy and the low-energy level-manifolds are distinct.  All of the levels approach zero for large $R$.  The lower panels show the spectra at $\mu=0.77$, the value appropriate to GaAs.  Here, levels from the upper and lower manifolds are clearly seen to cross and mix, once the inter-acceptor distance exceeds one effective Bohr radius.

\section{Explanation of Results}
\label{sec:explanation}
In order to better understand the degeneracy structure and level crossings observed in our results, the energy spectra were fit, as a function of $R$ and $\mu$, to an acceptor-acceptor Hubbard model derived in Appendix~\ref{app:Hubbard}.  That model depends on five parameters: $\epsilon$ (single-particle energy for holes with $F_z=\pm\frac{1}{2}$), $\epsilon'$ (single-particle energy for holes with $F_z=\pm\frac{3}{2}$), $t$ (hopping for holes with $F_z=\pm\frac{1}{2}$), $t'$ (hopping for holes with $F_z=\pm\frac{3}{2}$), and $U$ (on-site repulsion).  It yields sixteen energies, grouped in six distinct levels of known degeneracy.  The functional form of those energy levels is presented below, listed in an order that corresponds to the small-$\mu$ regime of Fig.~\ref{fig:energyR} (high energy to low).  In brackets are the $F_z^{\rm tot}$ labels appropriate to each level ($n$ labels denotes an $n$-fold degenerate level).
\begin{align}
E_{1} &= 2\epsilon & [0,\pm1] \nonumber\\
E_{5} &= \epsilon+\epsilon' + \frac{U}{2}\left[1-\sqrt{1+4\left(\frac{t-t'}{U}\right)^2}\right]  & [\pm1,\pm2]\nonumber\\
E_{2} &= 2\epsilon' & [0,\pm3]\nonumber\\
E_{4} &= 2\epsilon' + \frac{U}{2}\left[1-\sqrt{1+4\left(\frac{2t'}{U}\right)^2}\right] & [0]\nonumber\\
E_{6} &= \epsilon+\epsilon' + \frac{U}{2}\left[1-\sqrt{1+4\left(\frac{t+t'}{U}\right)^2}\right]  & [\pm1,\pm2]\nonumber\\
E_{3} &= 2\epsilon + \frac{U}{2}\left[1-\sqrt{1+4\left(\frac{2t}{U}\right)^2}\right] & [0]\nonumber\\
\label{eq:HubsimpleAll}
\end{align}
Note, first of all, that the degeneracy structure and $F_z^{\rm tot}$-labeling of the acceptor-acceptor Hubbard model is precisely that of our numerical Heitler-London results, at least in the not-too-large-$\mu$ and not-too-large-$R$ regime (see Fig.~\ref{fig:energyR}).  Both feature six distinct energy levels, of degeneracy 1, 4, 1, 3, 4, and 3 (from low energy to high), corresponding to $F_z^{\rm tot} = 0$, $F_z^{\rm tot} = \pm1,\pm2$, $F_z^{\rm tot} = 0$, $F_z^{\rm tot} = 0,\pm3$, $F_z^{\rm tot} = \pm1,\pm2$, and $F_z^{\rm tot} = 0,\pm1$.

This same degeneracy structure has been observed in calculations based on the Kohn-Luttinger framework \cite{sal16a,kav04,cli08,yak10,pas14}.  See Fig.~5 of Ref.~\onlinecite{sal16a}.  In that work, the two sets of one-fold and three-fold degenerate levels are viewed as singlets and triplets for heavy holes and light holes, while the four-fold degenerate levels denote states that mix heavy and light holes.

Note further, that in the limit that the primed ($F_z=\pm\frac{3}{2}$) parameters equal the unprimed ($F_z=\pm\frac{1}{2}$) parameters, our model spectrum coalesces to two levels of degeneracy 6 and 10, corresponding to the hydrogenic singlet and triplet, precisely as we see in our numerics for $\mu=0$.  Thus, it is clear that this simple model possesses many of the essential features required to explore our numerical results.

We therefore were able to fit those results to the expressions in Eq.~(\ref{eq:HubsimpleAll}) and thereby extracted values for the five model parameters as a function of $R$ and $\mu$.  Since the expressions are nonlinear functions of the parameters, we employed a nonlinear least-squares fit, as described in Appendix~\ref{app:fit}.

\begin{figure}
\centerline{\resizebox{3.25in}{!}{\includegraphics{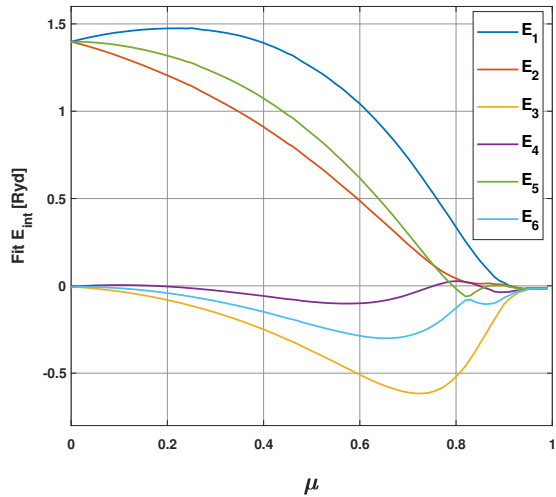}}}
\caption{Acceptor-acceptor Hubbard model energy spectra, obtained via fit to the numerical Heitler-London results shown in Fig.~\ref{fig:energyR}(a), plotted as a function of spin-orbit parameter $\mu$ for inter-acceptor distance $R=1a_B$.  The six curves correspond to the six expressions in Eq.~(\ref{eq:HubsimpleAll}), evaluated for the parameter values plotted in Fig.~\ref{fig:fitR1}.  Comparison with Fig.~\ref{fig:energyR}(a) defines the fidelity of the fit. }
\label{fig:fitData}
\end{figure}

\begin{figure}
\centerline{\resizebox{3.25in}{!}{\includegraphics{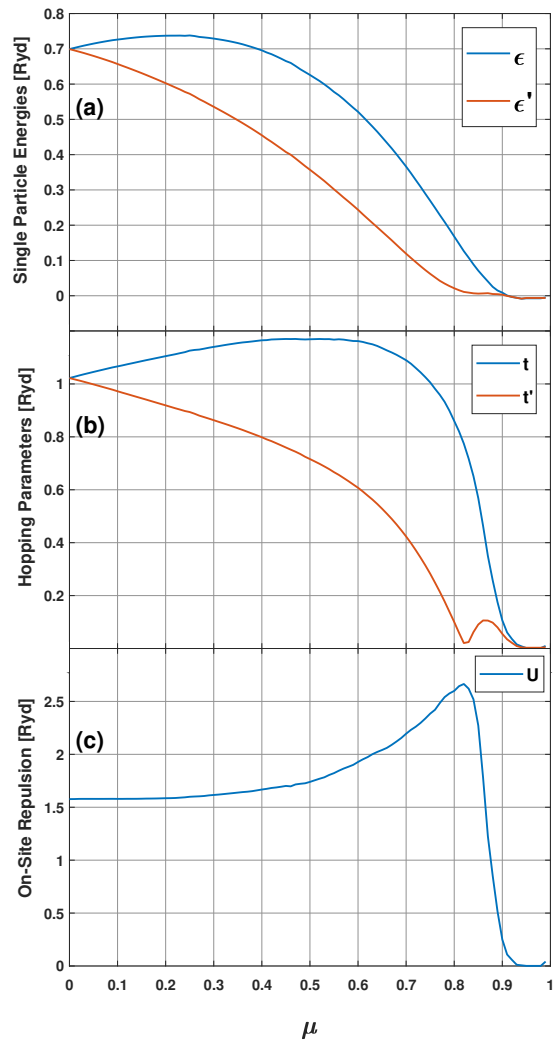}}}
\caption{Acceptor-acceptor Hubbard fit parameters plotted as a function of spin-orbit parameter $\mu$ for inter-acceptor distance $R=1a_B$.}
\label{fig:fitR1}
\end{figure}

\begin{figure}
\centerline{\resizebox{3.25in}{!}{\includegraphics{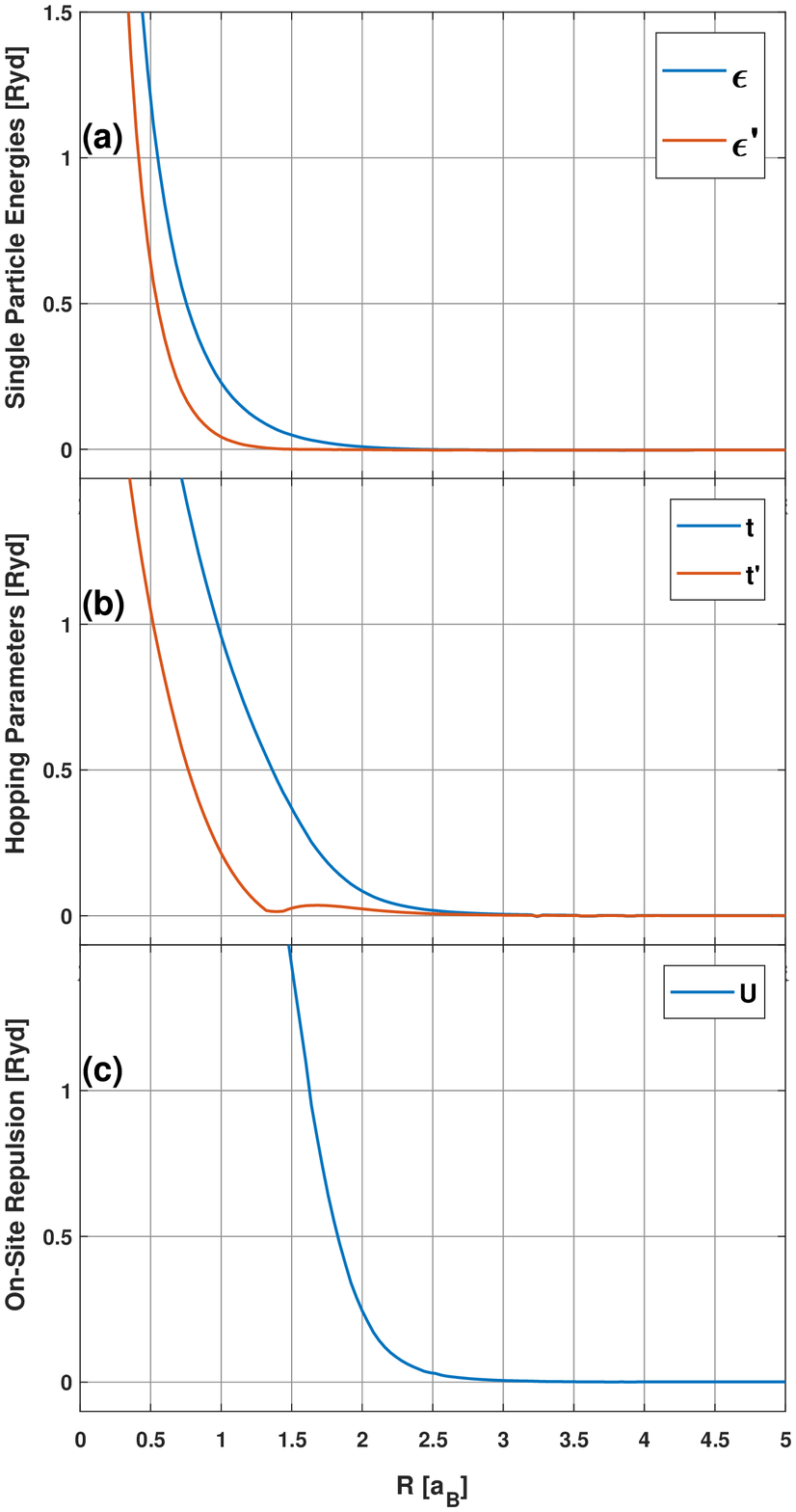}}}
\caption{Acceptor-acceptor Hubbard fit parameters plotted as a function of inter-acceptor distance $R$ for spin-orbit parameter $\mu=0.77$ (appropriate to GaAs).}
\label{fig:fitM77}
\end{figure}

The results of this fitting procedure are presented in Figs.~\ref{fig:fitData}-\ref{fig:fitM77}.  In Fig.~\ref{fig:fitData}, we plot the fit energy spectra as a function of $\mu$ for $R=1a_B$, to be compared with the numerical Heitler-London spectra plotted in Fig.~\ref{fig:energyR}(a).  The $\mu$-dependence of the extracted model parameters is plotted in Fig.~\ref{fig:fitR1}.  Note that the model is indeed capable of fitting our results in most instances.  At $\mu=0$, both numerics and fit display the two-level singlet-triplet spectrum of the hydrogen molecule.  The fit achieves this by equating primed and unprimed parameters.  As $\mu$ increases, the low-energy level splits into three levels of degeneracy 1, 4, and 1, and the high-energy level splits into three levels of degeneracy 3, 4, and 3.  This is achieved in the fit through a deviation of primed parameters from unprimed, with $\epsilon'<\epsilon$ and $t'<t$.  As $\mu$ increases further, the upper manifold of levels mixes with the lower manifold of levels, and just beyond $\mu=0.8$, the highest nondegenerate level meets the lowest 3-fold degenerate level and the two 4-fold degenerate levels cross.  In the Hubbard fit, this is a direct consequence of the parameter $t'$ reaching zero.  Looking at the form of Eq.~(\ref{eq:HubsimpleAll}), it is easy to see that setting $t'=0$ sets $E_2=E_4$ and $E_5=E_6$, creating precisely this meeting of levels.  Beyond this point, we either say that $t'$ goes negative and $E_6$ becomes greater than $E_5$, or we define $E_5$ to be greater than $E_6$ which keeps $t'$ positive, rebounding off of zero.  Since both 4-fold degenerate levels have the same $F_z^{\rm tot}$ labels, they are not distinguishable, so these two scenarios are entirely equivalent, just as there is no physical difference between a level crossing and a zero-gap anti-crossing.  For simplicity, we assume the latter.  As $\mu$ increases further, the level splittings diminish and all model parameters approach zero.

A few points worth noting:  Since we have fit the model expressions to our computed interaction energy, $E_{\rm int} = E - 2E_0$, and have included the $2/R$ acceptor-acceptor repulsion in that quantity, we have implicity added $1/R - E_0(\mu)$ to both $\epsilon$ and $\epsilon'$, where $E_0(\mu)$ is the Baldereschi-Lipari \cite{bal73} single-acceptor energy.  This redefines what we mean by these parameters (they are no longer single-hole energies), but does not affect the splitting between them.  Note also that in Fig.~\ref{fig:fitData}(a), there is a small range of $\mu$ just beyond 0.8 where the highest nondegenerate level exceeds the lowest 3-fold degenerate level.  Yet, the form of Eq.~(\ref{eq:HubsimpleAll}) prevents $E_4$ from exceeding $E_2$ for positive $U$ and real $t'$.  Thus, this inversion, seen in the Heitler-London numerics, is not captured by the Hubbard fit.  The best the fit can do is setting $E_4=E_2$, as is seen in Fig.~\ref{fig:fitData}(b).  Finally, the Heitler-London numerics reveal a small splitting of the 4-fold degenerate levels into two 2-fold degenerate levels for large $\mu$.  This splitting between $F_z^{\rm tot}=\pm1$ states and $F_z^{\rm tot}=\pm2$ states is not precluded by symmetry, but cannot be reproduced in the Hubbard fit, which is limited to six levels.

Results for larger inter-acceptor separation are qualitatively similar to the $R=1a_B$ case, with parameter magnitudes appropriately reduced.  Upper and lower level-manifolds mix at a smaller value of $\mu$, so $t'$ goes to zero sooner (for smaller $\mu$).  The $R$-dependence of the model parameters for fixed $\mu=0.77$ (the value appropriate to GaAs) is plotted in Fig.~\ref{fig:fitM77}.  As expected, hopping parameters decrease as inter-acceptor separation grows (as do $\epsilon$ and $\epsilon'$ due to the included $1/R$ term).  Note that there is an $R$ at which $t'$ hits zero, indicative of the level-crossings seen in the numerics.

The curve fits discussed above help to reveal a simplicity to the Heitler-London results that is difficult to grasp from the numerics alone.  As a function of increasing spin-orbit parameter, the system evolves from one much like the hydrogen molecule (a two-level spectrum with common parameters for $F_z=\pm\frac{1}{2}$ holes and $F_z=\pm\frac{3}{2}$ holes), to a six-level spectrum with splittings induced by the deviation of parameters for the two types of holes, to a thoroughly mixed spectrum with crossings induced by one of the hoppings going through zero.

\section{Conclusions}
\label{sec:conclusions}
In this investigation, we performed a Heitler-London analysis to study the nature of acceptor-acceptor interactions in $p$-doped semiconductors.  We began by developing an appropriate set of basis states, antisymmetric product states of two Baldereschi-Lipari \cite{bal73} single-acceptor wave functions, located on different sites.  Since the Baldereschi-Lipari states use an effective spin-$\frac{3}{2}$ model, there are four states per acceptor and therefore sixteen Heitler-London basis states.  Using these sixteen basis states, $16 \times 16$ Hamiltonian and overlap matrices were constructed, with matrix elements calculated numerically as a function of inter-acceptor distance, $R$, and spin-orbit parameter, $\mu$.  Solving the secular equation defined by these matrices, the acceptor-pair energy spectrum was computed as a function of $R$ and $\mu$.

For $\mu=0$, the sixteen energy eigenvalues are grouped into two levels, a 6-fold degenerate ground state corresponding to total effective spin even ($F^{\rm tot}=0,2$) and a 10-fold degenerate excited state corresponding to total effective spin odd ($F^{\rm tot}=1,3$).  This is simply the hydrogenic limit, where the acceptor wave functions reduce to hydrogen wave functions, and our Heitler-London calculation yields the familiar hydrogen-molecule singlet and triplet levels.

For nonzero $\mu$, total angular momentum $F^{\rm tot}$ is no longer a good quantum number, but if we let the quantization axis point along the line joining the two acceptors, $F_z^{\rm tot}$ still is.  The up-down symmetry of the acceptor-pair system requires that energy eigenvalues labeled by nonzero $F_z^{\rm tot}$ must be doubly degenerate ($+F_z^{\rm tot}$ states are degenerate with corresponding $-F_z^{\rm tot}$ states).  This reduces the maximum possible number of distinct energy levels from sixteen to ten: four with $F_z^{\rm tot}=0$, three with $F_z^{\rm tot}=\pm1$, two with $F_z^{\rm tot}=\pm2$, and one with $F_z^{\rm tot}=\pm3$.  Our calculation reveals that as $\mu$ increases from zero, the low-energy 6-fold degenerate level splits into three levels of (approximate) degeneracy 1, 4, and 1, and the high-energy 10-fold degenerate level splits into three levels of (approximate) degeneracy 3, 4, and 3, as shown in Fig.~\ref{fig:energyR}.

This degeneracy structure can be understood in terms of the acceptor-acceptor Hubbard model derived in Appendix~\ref{app:Hubbard} and fit to our Heitler-London results in Sec.~\ref{sec:explanation}.  Since there are four flavors of effective spin ($+\frac{3}{2}$, $+\frac{1}{2}$, $-\frac{1}{2}$, and $-\frac{3}{2}$), symmetry allows the single-acceptor energy and hopping parameters to take on different values for $\pm\frac{3}{2}$ versus $\pm\frac{1}{2}$, so the acceptor-acceptor Hubbard model provides five parameters: $\epsilon$, $\epsilon'$, $t$, $t'$, and $U$.  The splitting seen in our Heitler-London numerics for nonzero $\mu$ is nicely understood as a breaking of the equivalence between the primed and unprimed parameters of the Hubbard model.  In fact, this simple model precisely recovers the 1-4-1-3-4-3 degeneracy structure and $F_z^{\rm tot}$ labeling of our results.

For small-$\mu$, the low-energy manifold of levels remains distinct from the high-energy manifold of levels.  But as $\mu$ increases further, levels from lower and upper manifolds cross.  In terms of the acceptor-acceptor Hubbard model, such crossings are understood as the consequence of one of the hopping parameters, $t'$, passing through zero.  For large enough $\mu$, our Heitler-London results indicate an additional splitting of the 3-fold and 4-fold degenerate levels down to the 1-fold and 2-fold degeneracies required by symmetry, resulting in a complex ten-level energy spectrum.  This additional splitting appears to be beyond the simple Hubbard model.  As inter-acceptor distance increases, the $\mu$ value at which the lower and upper level manifolds start crossing and mixing gets smaller and smaller, shifting this interesting non-hydrogenic regime to smaller $\mu$.

Note that for silicon, a smaller-than-typical spin-orbit parameter and a larger-than-typical cubic parameter, means that corrections to the Baldereschi-Lipari single-acceptor wave functions \cite{bal73} will be larger than for most semiconductors.  Nevertheless, even for silicon, our results likely have the right qualitative behavior and the calculation herein provides a starting point from which corrections can be calculated.  To improve upon these results, one should include cubic corrections \cite{bal74} and account for the additional split-off bottom band when calculating the single-acceptor ground state wave functions.

It is clear from the present analysis that the energy spectrum of the acceptor-pair system can be far richer than the singlet-triplet spectrum of the hydrogen molecule or donor-pair system.  This distinction is significant and should therefore have interesting consequences for the properties of p-doped semiconductor systems.  Application to systems of acceptor-based qubits should provide a better understanding of, and potentially the ability to better control, two-qubit interactions.

Another system in which acceptor-acceptor interactions are particularly important is that of p-doped diluted magnetic semiconductors, like Ga$_{1-x}$Mn$_{x}$As. \cite{ohn98,mat98}  In such materials, where magnetic ions have been substitutionally inserted into III-V or II-VI semiconductors, acceptor-acceptor interactions can mediate the interactions between magnetic moments and thereby influence magnetic properties. \cite{wol96,dur02,zar02}  While such systems have been studied in detail \cite{fie03,fie05a,fie05b}, the complex nature of the acceptor-acceptor interactions is often neglected.  In future work, we intend to apply our acceptor-acceptor Heitler-London calculation directly to the case of Ga$_{1-x}$Mn$_{x}$As.  This will require some modifications to account for central cell corrections \cite{ben97,bha00,ber01}.

Our results also provide a basis for the development of a strong disorder renormalization group (SDRG) technique for studying the thermodynamic properties of insulating p-doped semiconductors, a generalization of the Bhatt-Lee SDRG scheme \cite{bha81,bha82}, but tailored to the complex nature of acceptor-pair interactions.  In future work, we plan to apply our acceptor-acceptor calculations toward this end, and see how well our model can quantitatively fit magnetic susceptibility data in Si:B \cite{roy86} and other p-doped semiconductors.

\begin{acknowledgments}
The authors are grateful to M. Miranda for providing access to the Hofstra Big Data Cluster and to B. Burrington and G. C. Levine for helpful discussions.  A.C.D. and K.E.C. were supported by funds provided by Hofstra University, including a Faculty Research and Development Grant (FRDG), a Presidential Research Award Program (PRAP) grant, and faculty startup funding.  R.N.B. was supported by DOE-BES Grant No.\ DE-SC0002140 and thanks the Aspen Center for Physics for their hospitality during the writing of this manuscript.
\end{acknowledgments}

\appendix
\section{Hubbard Model for the Acceptor-Acceptor Problem}
\label{app:Hubbard}
Within the Hubbard model, we represent each acceptor by a single orbital level centered at the location of that acceptor.  If the level is occupied by a hole of (effective) spin $F_z=\alpha$, the energy is $\epsilon_\alpha$.  The Pauli exclusion principle prohibits two holes of the same spin from occupying the same level.  If two holes of different spin, $\alpha$ and $\beta$, occupy the same level, the effect of hole-hole interactions is modeled by an on-site repulsion $U$, such that the energy of such a doubly-occupied level is taken to be $\epsilon_\alpha + \epsilon_\beta + U$.  For the acceptor-acceptor problem, we consider two acceptor sites, $A$ and $B$, to which a hole can be bound.  The single-hole Hamiltonian is expressed as $h = h_0 + \Delta h$, where $h_0$ is the single acceptor-hole Hamiltonian and $\Delta h$ is the attraction between the hole and the other acceptor. The single-hole matrix elements then take the form
\begin{eqnarray}
\langle A_\alpha | A_\beta \rangle &=& \langle B_\alpha | B_\beta \rangle = \delta_{\alpha \beta}\nonumber\\
\langle A_\alpha | B_\beta \rangle &=& \langle B_\alpha | A_\beta \rangle = s_\alpha \delta_{\alpha \beta}\nonumber\\
\langle A_\alpha |h| A_\beta \rangle &=& \langle B_\alpha |h| B_\beta \rangle = \epsilon_\alpha \delta_{\alpha \beta}\nonumber\\
\langle A_\alpha |h| B_\beta \rangle &=& \langle B_\alpha |h| A_\beta \rangle = t_\alpha \delta_{\alpha \beta}\nonumber\\
\label{eq:HubSingleMatElem}
\end{eqnarray}
where
\begin{eqnarray}
\langle A_\alpha |h_0| A_\beta \rangle &=& \langle B_\alpha |h_0| B_\beta \rangle = \epsilon_0 \delta_{\alpha \beta}\nonumber\\
\langle A_\alpha |\Delta h| A_\beta \rangle &=& \langle B_\alpha |\Delta h| B_\beta \rangle = \Delta \epsilon_\alpha \delta_{\alpha \beta}\nonumber\\
\epsilon_\alpha &\equiv& \epsilon_0 + \Delta \epsilon_\alpha .
\label{eq:HubEalpha}
\end{eqnarray}
The single-hole energies $\epsilon_\alpha$, hopping terms $t_\alpha$, and overlap terms $s_\alpha$ can, in general, be different for different $F_z=\alpha$, where the $z$-axis is defined along the line joining the acceptor sites.  Due to the cylindrical symmetry of the system, the definition of positive $z$ versus negative $z$ is arbitrary, thus these parameters can only depend on the absolute value of $\alpha$.  For a spin-$\frac{1}{2}$ system, like the familiar case of the hydrogen molecule Hubbard model \cite{ash76,alv02}, $\alpha=\{\frac{1}{2},-\frac{1}{2}\}$, so this restricts the problem to a single $\epsilon$ parameter, a single $t$ parameter, and a single $s$ parameter.  However, in the spin-$\frac{3}{2}$ system that we consider, $\alpha=\{\frac{3}{2},\frac{1}{2},-\frac{1}{2},-\frac{3}{2}\}$, so symmetry permits two flavors of each parameter, one for $\alpha=\pm\frac{1}{2}$ and another for $\alpha=\pm\frac{3}{2}$.  In what follows, we will let $\epsilon$, $t$, and $s$ refer to the parameters for $\alpha=\pm\frac{1}{2}$ and let $\epsilon'$, $t'$, and $s'$ refer to the parameters for $\alpha=\pm\frac{3}{2}$.  As we shall see, this difference goes a long way toward explaining the difference between the degeneracy structure of the acceptor-acceptor problem and that of the hydrogen molecule.

With the seven model parameters -- $\epsilon$, $\epsilon'$, $t$, $t'$, $s$, $s'$, and $U$ -- defined, we can proceed to evaluate the two-hole Hamiltonian in the basis of the antisymmetrized two-hole states.  There are two types of two-hole states to consider, states with one hole on each acceptor site
\begin{equation}
|A_\alpha B_\beta\rangle = \tfrac{1}{\sqrt{2}} \left(|A_\alpha(1) B_\beta(2)\rangle - |B_\beta(1) A_\alpha(2)\rangle \right)
\label{eq:HubbardAB}
\end{equation}
and states with two holes on the same acceptor site
\begin{equation}
|A_\alpha A_\beta\rangle = \tfrac{1}{\sqrt{2}} \left(|A_\alpha(1) A_\beta(2)\rangle - |A_\beta(1) A_\alpha(2)\rangle \right)
\label{eq:HubbardAA}
\end{equation}
\begin{equation}
|B_\alpha B_\beta\rangle = \tfrac{1}{\sqrt{2}} \left(|B_\alpha(1) B_\beta(2)\rangle - |B_\beta(1) B_\alpha(2)\rangle \right) .
\label{eq:HubbardBB}
\end{equation}
Note that states of the latter type are only unique for $\alpha > \beta$ since $|A_\alpha A_\alpha\rangle = 0$ and $|A_\beta A_\alpha\rangle = -|A_\alpha A_\beta\rangle$.  Thus if $\alpha$ can take $n$ values, there are $n^2$ $AB$-states, $(n^2-n)/2$ $AA$-states, and $(n^2-n)/2$ $BB$-states to consider.  In the spin-$\frac{1}{2}$ (hydrogen molecule) case \cite{ash76,alv02} where $n=2$, this means $4+1+1=6$ states in all.  But in the spin-$\frac{3}{2}$ case with which we are concerned, $n=4$ so there are 6 ways that $\alpha$ can be larger than $\beta$ and therefore $16+6+6=28$ states to consider.

The two-hole Hamiltonian, $H = h_1 + h_2 + V_{12}$, is the sum of the two single-hole Hamiltonians, $h_1$ and $h_2$, plus the hole-hole interaction term, $V_{12}$.  Within the Hubbard model, $V_{12}$ yields the on-site repulsion $U$ for diagonal matrix elements with both holes bound to the same acceptor, and is zero otherwise.  Making use of the definitions of the two-hole states [Eqs.~(\ref{eq:HubbardAB})-(\ref{eq:HubbardBB})] as well as the single-hole matrix elements [Eq.~(\ref{eq:HubSingleMatElem})], it is straightforward to compute the two-hole matrix elements.  The overlap matrix elements take the form
\begin{eqnarray}
\langle A_{\alpha'} B_{\beta'} | A_\alpha B_\beta \rangle &=& \delta_{\alpha' \alpha} \delta_{\beta' \beta} - s_{\alpha'}s_{\beta'} \delta_{\alpha' \beta} \delta_{\beta' \alpha}\nonumber\\
\langle A_{\alpha'} A_{\beta'} | A_\alpha B_\beta \rangle &=& s_{\beta'}\delta_{\alpha' \alpha} \delta_{\beta' \beta} - s_{\alpha'} \delta_{\alpha' \beta} \delta_{\beta' \alpha}\nonumber\\
\langle B_{\alpha'} B_{\beta'} | A_\alpha B_\beta \rangle &=& s_{\alpha'}\delta_{\alpha' \alpha} \delta_{\beta' \beta} - s_{\beta'} \delta_{\alpha' \beta} \delta_{\beta' \alpha}\nonumber\\
\langle A_{\alpha'} A_{\beta'} | A_\alpha A_\beta \rangle &=& \delta_{\alpha' \alpha} \delta_{\beta' \beta}\nonumber\\
\langle B_{\alpha'} B_{\beta'} | B_\alpha B_\beta \rangle &=& \delta_{\alpha' \alpha} \delta_{\beta' \beta}\nonumber\\
\langle A_{\alpha'} A_{\beta'} | B_\alpha B_\beta \rangle &=& s_{\alpha'}s_{\beta'}\delta_{\alpha' \alpha} \delta_{\beta' \beta}
\label{eq:HubDoubleElem}
\end{eqnarray}
and the Hamiltonian matrix elements take the form
\begin{eqnarray}
\langle A_{\alpha'} B_{\beta'} |H| A_\alpha B_\beta \rangle &=& (\epsilon_{\alpha'} + \epsilon_{\beta'})\delta_{\alpha' \alpha} \delta_{\beta' \beta} \nonumber\\ &-& (s_{\alpha'}t_{\beta'} + s_{\beta'}t_{\alpha'}) \delta_{\alpha' \beta} \delta_{\beta' \alpha}\nonumber\\
\langle A_{\alpha'} A_{\beta'} |H| A_\alpha B_\beta \rangle &=& (\epsilon_{\alpha'}s_{\beta'}+t_{\beta'})\delta_{\alpha' \alpha} \delta_{\beta' \beta} \nonumber\\ &-& (\epsilon_{\beta'}s_{\alpha'}+t_{\alpha'}) \delta_{\alpha' \beta} \delta_{\beta' \alpha}\nonumber\\
\langle B_{\alpha'} B_{\beta'} |H| A_\alpha B_\beta \rangle &=& (\epsilon_{\beta'}s_{\alpha'}+t_{\alpha'})\delta_{\alpha' \alpha} \delta_{\beta' \beta} \nonumber\\ &-& (\epsilon_{\alpha'}s_{\beta'}+t_{\beta'}) \delta_{\alpha' \beta} \delta_{\beta' \alpha}\nonumber\\
\langle A_{\alpha'} A_{\beta'} |H| A_\alpha A_\beta \rangle &=& (\epsilon_{\alpha'} + \epsilon_{\beta'} + U)\delta_{\alpha' \alpha} \delta_{\beta' \beta}\nonumber\\
\langle B_{\alpha'} B_{\beta'} |H| B_\alpha B_\beta \rangle &=& (\epsilon_{\alpha'} + \epsilon_{\beta'} + U)\delta_{\alpha' \alpha} \delta_{\beta' \beta}\nonumber\\
\langle A_{\alpha'} A_{\beta'} |H| B_\alpha B_\beta \rangle &=& (s_{\alpha'}t_{\beta'}+s_{\beta'}t_{\alpha'})\delta_{\alpha' \alpha} \delta_{\beta' \beta} . \nonumber\\
\label{eq:HubHamElem}
\end{eqnarray}
While the resulting Hamiltonian and overlap matrices, $\mathbf{H}$ and $\mathbf{S}$, are $28 \times 28$ matrices, most of those matrix elements are zero.  Note that each of the four $AB$-states with $\alpha=\beta$ only couples to itself.  Furthermore, the other 24 states can be organized into six sets of four states that only couple internally.  Each set is of the form $\{|A_\alpha B_\beta\rangle, |A_\beta B_\alpha\rangle, |A_\alpha A_\beta\rangle, |B_\alpha B_\beta\rangle\}$ for a particular $(\alpha,\beta)$ pair where $\alpha > \beta$.  The six such pairs are $(\alpha,\beta) = \{(\frac{3}{2},\frac{1}{2}), (\frac{3}{2},-\frac{1}{2}), (\frac{3}{2},-\frac{3}{2}), (\frac{1}{2},-\frac{1}{2}), (\frac{1}{2},-\frac{3}{2}), (-\frac{1}{2},-\frac{3}{2})\}$.  Thus, the $\mathbf{H}$ and $\mathbf{S}$ matrices are block diagonal with four $1 \times 1$ blocks and six $4 \times 4$ blocks.  So solving the $28 \times 28$ secular equation, $\det(\mathbf{H}-E\mathbf{S}) = 0$, reduces to the solution of four $1 \times 1$ secular equations and six $4 \times 4$ secular equations.

The four 1x1 blocks take the form
\begin{eqnarray}
S  &=& 1-s_\alpha^2\nonumber\\
H  &=& 2(\epsilon_\alpha - s_\alpha t_\alpha)\nonumber\\
\label{eq:Hub1x1Matrices}
\end{eqnarray}
and solving the secular equation yields two 2-fold degenerate energy levels
\begin{eqnarray}
E_1  &=& 2 \frac{\epsilon - st}{1-s^2}\nonumber\\
E_2  &=& 2 \frac{\epsilon' - s't'}{1-s'^2}
\label{eq:Hub1x1Energies}
\end{eqnarray}
where $E_1$ corresponds to $(\alpha,\beta) = \{(\frac{1}{2},\frac{1}{2}), (-\frac{1}{2},-\frac{1}{2})\}$ such that $F_z^{\rm tot} = \pm 1$ and $E_2$ corresponds to $(\alpha,\beta) = \{(\frac{3}{2},\frac{3}{2}), (-\frac{3}{2},-\frac{3}{2})\}$ such that $F_z^{\rm tot} = \pm 3$.

The six 4x4 blocks take the form
\begin{eqnarray}
S &=& \begin{bmatrix}  1  &  -s_\alpha s_\beta  &  s_\beta  &  s_\alpha  \\
                      -s_\alpha s_\beta  &   1  & -s_\alpha  & -s_\beta  \\
                       s_\beta  &  -s_\alpha  &  1 &  s_\alpha s_\beta  \\
                       s_\alpha  &  -s_\beta  &  s_\alpha s_\beta  & 1 \\
\end{bmatrix}\nonumber\\
H &=& \begin{bmatrix}  A  &  -B  &  C  &  D  \\
                      -B  &   A  & -D  & -C  \\
                       C  &  -D  & A+U &  B  \\
                       D  &  -C  &  B  & A+U \\
\end{bmatrix}
\label{eq:Hub4x4Matrices}
\end{eqnarray}
where
\begin{eqnarray}
A &=& \epsilon_\alpha + \epsilon_\beta \nonumber\\
B &=& t_\alpha s_\beta+t_\beta s_\alpha \nonumber\\
C &=& t_\beta + \epsilon_\alpha s_\beta \nonumber\\
D &=& t_\alpha + \epsilon_\beta s_\alpha
\label{eq:Hub4x4LettersABCD}
\end{eqnarray}
Solving the secular equation yields
\begin{equation}
E = \frac{K}{K^2-M^2}\left[ J-\frac{ML}{K}+\frac{U}{2} \left[ 1 \pm \sqrt{1+Q}\right]\right]
\label{eq:Hub4x4Energies}
\end{equation}
where
\begin{eqnarray}
Q &\equiv& 4\left(\frac{L}{U}\right)^2 +4\frac{M}{K^2U^2}\left(MJ(J+U)-KL(2J+U\right) \nonumber\\
J &\equiv& (\epsilon_\alpha +\epsilon_\beta) \pm (t_\alpha s_\beta + t_\beta s_\alpha) \nonumber\\
K &\equiv& 1 \pm (s_\beta s_\alpha) \nonumber\\
L &\equiv& (t_\alpha +\epsilon_\beta s_\alpha) \pm (t_\beta +\epsilon_\alpha s_\beta) \nonumber\\
M &\equiv& s_\alpha \pm s_\beta
\label{eq:Hub4x4LettersJKLM}
\end{eqnarray}
and the $\pm$ in the latter four equations of Eq.~(\ref{eq:Hub4x4LettersJKLM}) all must be chosen together.  The four eigenvalues emerge from the two choices of $\pm$ in Eq.~(\ref{eq:Hub4x4Energies}) and Eq.~(\ref{eq:Hub4x4LettersJKLM}).  If the Hubbard $U$ is taken to be large, then solutions that arise from the + in Eq.~(\ref{eq:Hub4x4Energies}) denote high-energy states that we shall discard.  Since there are two such states per $4 \times 4$ block and there are six such blocks, twelve high-energy levels may be discarded in this manner.  The sixteen energy levels that remain are those appropriate for comparison with the results of our Heitler-London analysis (which neglected the high-energy levels from the start).

For the two $(\alpha,\beta)$ pairs where $\beta=-\alpha$, ($\frac{1}{2}$,$-\frac{1}{2}$) and ($\frac{3}{2}$,$-\frac{3}{2}$), Eq.~(\ref{eq:Hub4x4LettersJKLM}) simplifies nicely when the minus sign is chosen. In these cases
\begin{eqnarray}
Q &=& 0 \nonumber\\
J &=& 2(\epsilon_\alpha - t_\alpha s_\alpha) \nonumber\\
K &=& 1 - s_\alpha^2 \nonumber\\
L &=& 0 \nonumber\\
M &=& 0
\label{eq:Hub4x4LettersJKLM2}
\end{eqnarray}
and $E=\frac{J}{K}$, which yields precisely the same energies as we already obtained in Eq.~(\ref{eq:Hub1x1Energies}) from the $1 \times 1$ blocks, $E_1$ for ($\frac{1}{2}$,$-\frac{1}{2}$) and $E_2$ for ($\frac{3}{2}$,$-\frac{3}{2}$). Both levels are therefore triply degenerate, corresponding to $F_z^{\rm tot} = 0,\pm 1$ and $F_z^{\rm tot} = 0,\pm 3$ respectively.  When the plus sign in Eq.~(\ref{eq:Hub4x4LettersJKLM}) is chosen instead, these two pairs yield two distinct, nondegenerate energy levels, both corresponding to $F_z^{\rm tot} = 0$.

Because $s_\alpha$, $t_\alpha$, and $\epsilon_\alpha$ only depend on the absolute value of $\alpha$, the energy level solutions from the pair ($\frac{3}{2}$,$\frac{1}{2}$) must be the same as those from ($\frac{3}{2}$,$-\frac{1}{2}$), and the solutions for ($\frac{1}{2}$,$-\frac{3}{2}$) must equal those for ($-\frac{1}{2}$,$-\frac{3}{2}$).  There can only then be four more unique levels at most, however we do not get this many.  These four pairs all have one $\pm\frac{1}{2}$ and one $\pm\frac{3}{2}$.  If the plus sign in Eq.~(\ref{eq:Hub4x4LettersJKLM}) is chosen, then $J$, $K$, $L$, and $M$ do not depend on the order that the $\pm\frac{1}{2}$ and $\pm\frac{3}{2}$ appear, thus neither does $E$.  If the minus is chosen, then $J$ and $K$ are again independent of the order, but $L$ and $M$ change sign depending on which is first.  Looking at $Q$ in Eq.~(\ref{eq:Hub4x4LettersJKLM}) and $E$ in Eq.~(\ref{eq:Hub4x4Energies}), each $L$ and $M$ always multiply or divide another $L$ or $M$, cancelling out any effect that this sign change could yield.  Therefore, there are only two 4-fold degenerate energy levels that arise from these four $(\alpha,\beta)$ pairs, one given by the plus sign in Eq.~(\ref{eq:Hub4x4LettersJKLM}) and the other given by the minus sign.  Both correspond to $F_z^{\rm tot} = \pm 1, \pm 2$.  A quick glance at Fig.~\ref{fig:energyR} reveals that this degeneracy structure and $F_z^{\rm tot}$-labeling is precisely that of our Heitler-London results!

The above expressions simplify further in the limit that we can neglect the overlap parameters ($s \approx s' \approx 0$).  Then Eq.~(\ref{eq:Hub1x1Energies}) simply becomes
\begin{equation}
E_{1} = 2\epsilon
\label{eq:HubsimpleE1}
\end{equation}
for ($\frac{1}{2}$,$\frac{1}{2}$) or ($-\frac{1}{2}$,$-\frac{1}{2}$) and
\begin{equation}
E_{2} = 2\epsilon'
\label{eq:HubsimpleE2}
\end{equation}
for ($\frac{3}{2}$,$\frac{3}{2}$) or ($-\frac{3}{2}$,$-\frac{3}{2}$).  Furthermore, Eqs.~(\ref{eq:Hub4x4Energies}) and (\ref{eq:Hub4x4LettersJKLM}) simplify to
\begin{equation}
E = \epsilon_\alpha +\epsilon_\beta + \frac{U}{2}\left[1-\sqrt{1+4\left(\frac{t_\alpha\pm t_\beta}{U}\right)^2}\right].
\label{eq:Hub4x4simple}
\end{equation}
For ($\frac{1}{2}$,$-\frac{1}{2}$), Eq.~(\ref{eq:Hub4x4simple}) reduces to $E_1$ if the minus sign is chosen and to
\begin{equation}
E_{3} = 2\epsilon + \frac{U}{2}\left[1-\sqrt{1+4\left(\frac{2t}{U}\right)^2}\right]
\label{eq:HubsimpleE3}
\end{equation}
if the plus sign is chosen.  Similarly, for ($\frac{3}{2}$,$-\frac{3}{2}$), it reduces to $E_2$ with the minus sign and
\begin{equation}
E_{4} = 2\epsilon' + \frac{U}{2}\left[1-\sqrt{1+4\left(\frac{2t'}{U}\right)^2}\right]
\label{eq:HubsimpleE4}
\end{equation}
with the plus sign.  For the other four pairs, ($\frac{3}{2}$,$\frac{1}{2}$), ($\frac{3}{2}$,$-\frac{1}{2}$), ($\frac{1}{2}$,$-\frac{3}{2}$) and ($-\frac{1}{2}$,$-\frac{3}{2}$), the minus sign yields
\begin{equation}
E_{5} = \epsilon+\epsilon' + \frac{U}{2}\left[1-\sqrt{1+4\left(\frac{t-t'}{U}\right)^2}\right]
\label{eq:HubsimpleE5}
\end{equation}
and the plus sign yields
\begin{equation}
E_{6} = \epsilon+\epsilon' + \frac{U}{2}\left[1-\sqrt{1+4\left(\frac{t+t'}{U}\right)^2}\right] .
\label{eq:HubsimpleE6}
\end{equation}
These six energy levels, of degeneracy 3, 3, 1, 1, 4, and 4 respectively, and expressed as a function of the five parameters, $\epsilon$, $\epsilon'$, $t$, $t'$, and $U$, define the spectrum of the acceptor-acceptor Hubbard model.  In Sec.~\ref{sec:explanation}, we fit this model to the results of our acceptor-acceptor Heitler-London calculation in order to better understand the nature of those results.

\section{Nonlinear Least-Squares Fit}
\label{app:fit}
A least-squares fitting procedure was used to fit our numerical Heitler-London results to the acceptor-acceptor Hubbard model expressions in Eq.~(\ref{eq:HubsimpleAll}) and thereby extract values for the five parameters of the model ($\epsilon$, $\epsilon'$, $t$, $t'$, and $U$) as a function of $R$ and $\mu$.  Since those expressions depend on the parameters in a nonlinear manner, we performed a nonlinear least-squares fit via an iterative procedure.

For each value of $R$ and $\mu$, we begin with an initialization algorithm that generates an initial guess for the five parameters.  From the numerical results, we extract values for $E_1$, $E_2$, $E_3$, $E_4$, $E_5$, and $E_6$.  We then set $\epsilon$ and $\epsilon'$ via the expressions for $E_1$ and $E_2$ in Eq.~(\ref{eq:HubsimpleAll}).  A sequence of $U$ values are then considered from 0 through 5 in steps of 0.001.  For each $U$, we set $t$ and $t'$ via the expressions for $E_3$ and $E_4$ and compute the resulting values of $E_5$ and $E_6$.  Minimization of the error in $E_5$ and $E_6$ yields an optimal initial guess for the parameters.

With the initial guess in hand, we linearize the nonlinear expressions in Eq.~(\ref{eq:HubsimpleAll}) about these initial parameter values, resulting in linearized expressions of the form
\begin{equation}
E_i - E_{0i} = \sum_{j=1}^5 \frac{\partial E_{i}}{\partial x_j}\Big|_0 (x_j - x_{0j})
\label{eq:FitLinearized}
\end{equation}
where the $E_i$ are the energies to be fit, the $E_{0i}$ are the model expressions evaluated at the initial parameter values, the $\frac{\partial E_{i}}{\partial x_j}\big|_0$ are partial derivatives of the model expressions with respect the model parameters (evaluated at the initial parameter values), the $x_j$ are the parameters to be determined, and the $x_{0j}$ are the initial parameter values. Although there are six expressions in Eq.~(\ref{eq:HubsimpleAll}), the degeneracies yield sixteen equations in all, to be used to fit all sixteen energies in the spectra.  The above therefore defines a matrix equation of the form
\begin{equation}
\mathbf{y} = \mathbf{M} \mathbf{x}
\label{eq:FitLinearizedMatrix}
\end{equation}
where $\mathbf{y}$ is a 16-vector of energy differences, $\mathbf{M}$ is a $16 \times 5$ matrix of partials, and $\mathbf{x}$ is a 5-vector of parameter differences to be determined.  This system of linear equations is overdetermined and therefore amenable to least-squares analysis.  The linear least-squares estimate of its solution is given by
\begin{equation}
\mathbf{x} = (\mathbf{M}^{\rm T}\mathbf{M})^{-1}\mathbf{M}^{\rm T} \mathbf{y}
\label{eq:FitInverse}
\end{equation}
where $(\mathbf{M}^{\rm T}\mathbf{M})^{-1}\mathbf{M}^{\rm T}$ is the pseudo-inverse of the rectangular matrix $\mathbf{M}$. \cite{gel74}  From the computed $\mathbf{x}$, we extract an updated estimate of the five model parameters, $x_j$, which now replace our initial guess as the new $x_{0j}$ for the next iteration.  This process is then repeated (up to thirty times) until further iteration yields negligible corrections.  The entire procedure is repeated for each value of $R$ and $\mu$, producing best-fit parameter values and energy spectrum estimates as a function of $R$ and $\mu$.  The insight thereby gleaned is discussed in Sec.~\ref{sec:explanation}.

\end{document}